\documentclass[letter,11pt]{article}
\pdfoutput=1
\usepackage{jheppub}
\usepackage[dvipsnames,table,xcdraw]{xcolor}
\usepackage[T1]{fontenc}
\usepackage[colorlinks=true, linkcolor=blue, urlcolor=blue, citecolor=blue, anchorcolor=blue]{hyperref}
\usepackage{subcaption}

\title{Gradient and Hessian-Based Temperature Estimator in Lattice Gauge Theories: A Diagnostic Tool for Stability and Consistency in Numerical Simulations}

\author[a,b]{Navdeep Singh Dhindsa,}
\author[c]{Anosh Joseph,}
\author[d]{Vamika Longia}

\affiliation[a]{The Institute of Mathematical Sciences, CIT Campus, Chennai, 600113, India}
\affiliation[b]{Department of Theoretical Physics, Tata Institute of Fundamental Research, \\ Homi Bhabha Road, Mumbai 400005, India}
\affiliation[c]{National Institute for Theoretical and Computational Sciences, \\ School of Physics, and Mandelstam Institute for Theoretical Physics,\\ University of the Witwatersrand, Johannesburg, Wits 2050, South Africa}
\affiliation[d]{Department of Physical Sciences, Indian Institute of Science Education and Research - Mohali, Knowledge City, Sector 81, SAS Nagar, Punjab 140306, India}

\emailAdd{navdeep@theory.tifr.res.in}
\emailAdd{anosh.joseph@wits.ac.za}
\emailAdd{vamika.longia@gmail.com}

\abstract{We present a field configuration-based temperature estimator in lattice gauge theories, constructed from the gradient and Hessian of the Euclidean action. 
Adapted from geometric formulations of entropy in classical statistical mechanics, this estimator provides a gauge-invariant, non-kinetic diagnostic of thermodynamic consistency in Monte Carlo simulations. 
We validate the method in compact U(1) lattice gauge theories across one, two, and four dimensions, comparing the estimated configurational temperature with the conventional temperature set by the temporal extent of the lattice. 
Our results show that the estimator accurately reproduces the input temperature and remains robust across a range of lattice volumes and coupling strengths. 
The temperature estimator offers a general-purpose diagnostic for lattice field theory simulations, with potential applications to non-Abelian theories, anisotropic lattices, and real-time monitoring in hybrid Monte Carlo algorithms.
}

\begin{document}
\maketitle
\flushbottom

\section{Introduction}
\label{sec:intro}

In molecular dynamics simulations, we can use conserved quantities such as energy and momentum as effective indicators of algorithmic correctness. 
Although the conservation of these quantities does not guarantee a simulation's validity, violations typically signal programming or numerical errors, which are often easy to detect.

In contrast, canonical Monte Carlo simulations lack straightforward conservation laws that can be used for validation. 
In particular, there has traditionally been no method for directly computing the thermodynamic temperature using only the field configurational data. 
As a result, algorithm validation in the canonical ensemble relies largely on comparison with known thermodynamic properties.
This approach can become problematic when studying novel systems or exploring state points for which no reference data exist.

A significant advance in this area was the work of Rugh, who derived an expression for temperature based on a geometric analysis of phase space within the microcanonical ensemble \cite{PhysRevLett.78.772}. 
His formulation relates the temperature to the curvature of the constant-energy hypersurface and offers a purely dynamical definition grounded in the phase space structure.

Building on Rugh's insight, Butler {\it et al.} proposed a configurational definition of temperature applicable to canonical ensembles \cite{10.1063/1.477301}. 
This expression requires only configurational information, namely, gradients and curvatures of the potential energy or action, and is therefore ideally suited to Monte Carlo simulations, which do not sample momenta. 
Interestingly, this type of definition, based on mean-square forces, was not entirely new: related ideas can be traced back to earlier works, including the 1952 Russian-language edition of {\it Statistical Physics} by Landau and Lifshitz \cite{Landau1952}, and even more indirectly to Tolman's 1938 textbook {\it The Principles of Statistical Mechanics}, which discusses generalized equipartition without explicitly formulating the concept \cite{Tolman1979}. 
A later historical account by Hoover also notes this lineage \cite{Hoover2007}.

In Ref. \cite{PhysRevE.62.4757}, the authors demonstrate that arbitrary phase-space vector fields can be used to construct phase functions whose ensemble averages yield the thermodynamic temperature. They establish conditions under which these functions are valid in systems with periodic boundaries and within the molecular dynamics (MD) ensemble, supporting their results with simulations involving short-ranged potentials. In Ref. \cite{10.1063/1.477301}, the configurational temperature is shown, via Lennard-Jones simulations, to respond rapidly and accurately to changes in the input temperature, even when the system deviates from global thermodynamic equilibrium. A numerical test of the temperature estimator in the canonical ensemble is presented in Ref. \cite{PhysRevE.94.062113}, where the authors simulate the two-dimensional XY model using a generalized Wolff cluster algorithm. Broader discussions and generalizations of the underlying theoretical framework can be found in Refs. \cite{1998JPhA...31.7761R, PhysRevE.62.4757, 10.1063/1.1348024}, while further applications and validation in MD simulations are provided in Refs. \cite{10.1063/1.477301, 10.1063/1.480995}.

Although the configurational temperature was initially proposed as a tool for thermostat design and system control in numerical simulations, Butler {\it et al.} emphasized its practical value as a diagnostic.
They demonstrated its utility in detecting sampling inconsistencies and numerical errors in Monte Carlo simulations. 
In modern simulations, the configurational temperature can provide a means of assessing thermodynamic consistency independently of the microscopic momentum variables associated with kinetic definitions of the temperature. 

Understanding thermodynamic properties in lattice field theory simulations requires precise control over temperature. 
In such simulations, the physical temperature is typically introduced through the temporal lattice extent or inverse coupling.
However, discretization artifacts can lead to significant deviations in physical observables, such as a shift in the transition temperature at finite lattice spacing. 
These effects vanish only in the continuum limit; thus, verifying that the simulation temperature matches the intended physical temperature is a nontrivial task.

The microscopic temperature estimator offers a nontrivial and gauge-invariant way to verify whether a system is truly sampling from the expected thermal distribution.
It is especially useful in lattice gauge theories where discretization and algorithmic artifacts can shift critical temperatures or other thermodynamic properties. 
Using this estimator, we can independently measure the temperature and compare it with the input parameter to diagnose such issues.

The structure of the paper is as follows. 
In Section~\ref{sec:tempest}, we introduce the concept of a microscopic temperature estimator, based on Refs. \cite{PhysRevLett.78.772, 10.1063/1.477301}, and outline its theoretical basis and practical motivation. 
This section sets the stage for using the estimator as a diagnostic tool in equilibrium simulations. 
In Section~\ref{sec:u1}, we apply the estimator to compact U(1) lattice gauge theories in various dimensions, where analytic or high-precision numerical results are available. 
We explore how the measured temperature compares to the input temperature.
Finally, in Section~\ref{sec:conc}, we summarize our findings and discuss possible extensions of this approach to more complex theories.

\section{Configurational Temperature Estimator}
\label{sec:tempest}

This section outlines the construction of the microscopic temperature estimator entirely from the sampled field configurations.
In Hybrid Monte Carlo (HMC) simulations, where the momenta are auxiliary, this estimator can provide a nontrivial check on thermodynamic consistency. 

\subsection{Derivation of the configurational temperature estimator}

From elementary thermodynamics, we have the first law expressed as
\begin{equation}
dE = T dS - p dV.
\end{equation}
For an isochoric transformation, the rate of change of entropy with an increase in energy determines the inverse temperature:
\begin{equation}
\frac{1}{T} = \left( \frac{\partial S}{\partial E} \right)_V.
\label{eq:dE}
\end{equation}

Let $\vec{\Gamma}$ denote a point in phase space, comprising all position and momentum coordinates of the system
\begin{equation}
\vec{\Gamma} \equiv \left( q_1, q_2, \ldots, q_N; p_1, p_2, \ldots, p_N \right),
\end{equation}
where $q_i$ and $p_i$ represent the position and momentum of the $i$-th particle, respectively.

We define a microcanonical ensemble corresponding to a Hamiltonian $H(\vec{\Gamma})$ as the set of phase space points for which the total energy lies within a narrow interval $[E, E + \Delta E]$, with $\Delta E \ll E$. 
Under the assumption of equal {\it a priori} probabilities, the entropy associated with this ensemble is proportional to the volume of this energy shell in phase space:
\begin{align}
S(E) &= k_B \ln \Omega_\Gamma(E, N, V) \nonumber \\
&= k_B \ln \left( \int_{\mu C(E)} d\vec{\Gamma} \right),
\end{align}
where $\mu C(E)$ denotes the microcanonical shell, defined as the set of all $\vec{\Gamma}$ satisfying $E \leq H(\vec{\Gamma}) \leq E + \Delta E$. 
To compute the derivative $\partial S/ \partial E$, we examine how this phase-space volume changes under an infinitesimal shift in energy. 
Specifically, we introduce a displacement of each phase space point along a vector field $\vec{n}(\vec{\Gamma})$ that is constructed to increase the energy uniformly across the ensemble. 
This procedure amounts to mapping the set $\mu C(E)$ to a nearby energy shell $\mu C(E + \Delta E)$ so that the displacement induces a constant energy increase $\Delta E$ to leading order.

This construction allows us to evaluate the change in entropy purely from the geometry of the phase space, ultimately leading to an expression for the temperature in terms of configurational derivatives of the Hamiltonian with respect to the underlying coordinates.

To evaluate the change in entropy with respect to energy, we consider an infinitesimal transformation that shifts the system's energy from $E$ to $E + \Delta E$ by displacing phase space points along a specific direction in configuration space. 
The displaced point is defined by
\begin{equation}
\vec{\Gamma}'(\vec{\Gamma}) = \vec{\Gamma} + \Delta E ~ \frac{\vec{\nabla}_{\vec{q}} H(\vec{\Gamma})}{\vec{\nabla}_{\vec{q}} H(\vec{\Gamma}) \cdot \vec{\nabla}_{\vec{q}} H(\vec{\Gamma})},
\end{equation}
where
\begin{equation}
\vec{\nabla}_{\vec{q}} = \left( \frac{\partial}{\partial q_1}, \dots, \frac{\partial}{\partial q_N} \right),
\end{equation}
is the configurational gradient operator. 
This transformation moves each point in configuration space along the direction of increasing energy, ensuring a uniform increase in energy by $\Delta E$. 
Only the configurational components of the Hamiltonian gradient are considered, in keeping with the goal of expressing thermodynamic quantities in terms of positional degrees of freedom alone.

The vector field along which we displace the configurations is given by
\begin{equation}
\vec{n}(\vec{\Gamma}) \equiv \frac{\vec{\nabla}_{\vec{q}} H(\vec{\Gamma})}{\vec{\nabla}_{\vec{q}} H(\vec{\Gamma}) \cdot \vec{\nabla}_{\vec{q}} H(\vec{\Gamma})}.
\label{eq:nvector_Gamma}
\end{equation}
This choice ensures that the displacement yields a change in the Hamiltonian of exactly $\Delta E$, to leading order. 

To see this, consider the analogy of a topographic landscape where the height corresponds to the energy $H(\vec{q})$ of a configuration $\vec{q}$. 
Contour lines of constant energy represent level surfaces in configuration space. 
Moving tangentially to the contour leaves the energy unchanged, while motion in the direction of the gradient increases the energy most rapidly. 
The gradient $\vec{\nabla}_{\vec{q}} H$ therefore points normal to the contour and indicates both the direction and rate of steepest energy ascent.

We now consider a small displacement of the configuration vector:
\begin{equation}
\vec{q}~' = \vec{q} + \epsilon ~ \vec{n}(\vec{\Gamma}),
\end{equation}
with $\epsilon = \Delta E$ and $\vec{n}(\vec{\Gamma})$ as defined in Eq.~\eqref{eq:nvector_Gamma}. 
The normalization by $\vec{\nabla}_{\vec{q}} H \cdot \vec{\nabla}_{\vec{q}} H$ ensures that the change in energy is precisely $\Delta E$. 
Expanding the Hamiltonian to leading order, using a Taylor series, yields
\begin{equation}
H(\vec{q}~') = H(\vec{q}) + \vec{\nabla}_{\vec{q}} H \cdot (\vec{q}~' - \vec{q}) + \mathcal{O}((\vec{q}~' - \vec{q})^2).
\end{equation}
Substituting the displacement vector
\begin{equation}
\vec{q}~' - \vec{q} = \Delta E ~ \vec{n}(\vec{\Gamma}),
\end{equation}
we obtain
\begin{align}
H(\vec{q}~') &= H(\vec{q}) + \Delta E ~ \vec{\nabla}_{\vec{q}} H \cdot \vec{n}(\vec{\Gamma}) + \mathcal{O}((\Delta E)^2) \nonumber \\
&= H(\vec{q}) + \Delta E  + \mathcal{O}((\Delta E)^2),
\end{align}
which confirms that the Hamiltonian increases by exactly $\Delta E$ to linear order in the displacement. 
This controlled shift in energy forms the basis for computing thermodynamic derivatives such as $\partial S / \partial E$ in a purely configurational framework.

To leading order in $\Delta E$, the displacement of phase space points described above results in a uniform change in energy, independent of the initial phase space vector $\vec{\Gamma} \in \mu C(E)$. 
That is, all points on the constant-energy hypersurface are mapped to a nearby surface with energy $E + \Delta E + \mathcal{O}((\Delta E)^2)$.
We can compute the corresponding change in entropy by evaluating the transformation of the phase space volume under this displacement. 
Given the map $\vec{q} \mapsto \vec{q}~'$, the entropy of the displaced ensemble is
\begin{align}
S(E + \Delta E) &= k_B \ln \int_{\mu C(E + \Delta E)} d\vec{\Gamma} \nonumber \\
&= k_B \ln \int_{\mu C(E)} \left| \frac{\partial \vec{\Gamma}'}{\partial \vec{\Gamma}} \right| d\vec{\Gamma}.
\end{align}
The integrand includes the Jacobian determinant associated with the phase space transformation.

Since only the positions $\vec{q}$ are displaced while momenta remain fixed, the Jacobian is determined by the configuration space transformation:
\begin{equation}
J(\vec{q}) = \left| \frac{\partial \vec{q}~'}{\partial \vec{q}} \right| = 1 + \Delta E ~ \vec{\nabla}_{\vec{q}} \cdot \vec{n}(\vec{q}),
\end{equation}
where $\vec{n}(\vec{q})$ is the normalized displacement vector defined earlier. 
For completeness, the full phase space Jacobian may also be written as
\begin{equation}
J(\vec{\Gamma}) = \left| \frac{\partial \vec{\Gamma}'(\vec{\Gamma})}{\partial \vec{\Gamma}} \right| = 1 + \Delta E ~ \vec{\nabla}_{\vec{q}} \cdot \vec{n}(\vec{\Gamma}).
\label{eq:J_Gamma}
\end{equation}

This Jacobian captures the local volume change in phase space, and its logarithm contributes to the change in entropy under the transformation, as used below
\begin{align}
\Delta S &= S(E + \Delta E) - S(E) \nonumber \\
&= k_B \ln \left( 1 + \Delta E ~ \vec{\nabla}_{\vec{q}} \cdot \vec{n}(\vec{q}) \right) \nonumber \\
&\approx k_B \Delta E ~ \vec{\nabla}_{\vec{q}} \cdot \vec{n}(\vec{q}),
\end{align}
where we have used the approximation $\ln(1 + \delta) \approx \delta$ valid for small $\delta$.

Taking the limit as $\Delta E \to 0$, we obtain an expression for the derivative of entropy with respect to energy:
\begin{equation}
\frac{\partial S}{\partial E} = \lim_{\Delta E \to 0} \frac{\Delta S}{\Delta E} = k_B ~ \left\langle \vec{\nabla}_{\vec{q}} \cdot \vec{n}(\vec{q}) \right\rangle,
\end{equation}
where the average is taken over the ensemble.
Substituting the definition of $\vec{n}(\vec{q})$, this yields
\begin{equation}
\frac{1}{k_B T} = \frac{1}{k_B} \frac{\partial S}{\partial E} = \left\langle \vec{\nabla}_{\vec{q}} \cdot \left( \frac{\vec{\nabla}_{\vec{q}} H}{\vec{\nabla}_{\vec{q}} H \cdot \vec{\nabla}_{\vec{q}} H} \right) \right\rangle.
\end{equation}

Only the configurational contribution to the Hamiltonian is relevant in many canonical ensemble simulations, particularly Monte Carlo algorithms that omit momentum variables. 
Assuming $H(\vec{q}) = \Phi(\vec{q})$, where $\Phi$ is the potential energy (or the action), the expression reduces to
\begin{equation}
\frac{1}{k_B T} = \left\langle \vec{\nabla}_{\vec{q}} \cdot \left( \frac{\vec{\nabla}_{\vec{q}} \Phi}{\vec{\nabla}_{\vec{q}} \Phi \cdot \vec{\nabla}_{\vec{q}} \Phi} \right) \right\rangle + \mathcal{O}\left( \frac{1}{N} \right).
\label{eq:temp_eq}
\end{equation}

By the equivalence of ensembles, Eq.~\eqref{eq:temp_eq} remains valid in the canonical ensemble in the thermodynamic limit $N \to \infty$. 
This configurational definition of temperature thus provides a powerful diagnostic for validating sampling consistency in simulations that operate without explicit momentum variables\footnote{This formalism is not suited for systems where the momentum variables play a crucial role. For example, in Hamiltonian lattice gauge theory formalism, as opposed to Euclidean path integral formalism, the gauge field is the coordinate, the electric field is the canonical momentum, and both are physical observables.}.

We can rewrite the expression in Eq. \eqref{eq:temp_eq} in terms of the Hessian matrix and the gradient vector of $\Phi(\vec{q})$.
It helps express the configurational temperature in a more compact and numerically useful way.

Let us denote the gradient and Hessian in the following way:
\begin{equation}
\vec{g} = \vec{\nabla}_{\vec{q}} \Phi(\vec{q}), ~~ g \in {\mathbb R}^N
\end{equation}
and
\begin{equation}
{\mathbb H} = \vec{\nabla}_{\vec{q}} \vec{\nabla}_{\vec{q}}^T \Phi(\vec{q}), ~~ {\mathbb H} \in {\mathbb R}^{N \times N}.
\end{equation}

Then, the quantity
\begin{equation}
\vec{\nabla}_{\vec{q}} \cdot \left( \frac{\vec{\nabla}_{\vec{q}} \Phi}{\vec{\nabla}_{\vec{q}} \Phi \cdot \vec{\nabla}_{\vec{q}} \Phi} \right)
\end{equation}
can be written using the product rule (divergence of a vector field), as
\begin{equation}
\vec{\nabla}_{\vec{q}} \cdot \left( \frac{\vec{g}}{\vec{g} \cdot \vec{g}} \right) = \frac{1}{\vec{g} \cdot \vec{g}} \vec{\nabla}_{\vec{q}} \cdot \vec{g} - \frac{2}{\left( \vec{g} \cdot \vec{g} \right)^2} \sum_{i, j} g_i H_{ij} g_j.
\label{eq:another-form}
\end{equation}

This gives us
\begin{equation}
\vec{\nabla}_{\vec{q}} \cdot \left( \frac{\vec{g}}{\vec{g} \cdot \vec{g}} \right) = \frac{{\rm Tr} ({\mathbb H})}{| \vec{g} |^2} - 2 \frac{\vec{g}^T {\mathbb H} \vec{g}}{|\vec{g}|^4}.
\end{equation}

In higher dimensions, configurations sampled via $e^{-\Phi}$ tend to be smoother due to averaging over more directions, and the number of degrees of freedom increases. 
As a result, the gradient vector accumulates contributions from many local directions, leading to weaker directional alignment with the curvature. 
Consequently, the directional term $\vec{g}^{T} \mathbb{H} \vec{g} / |\vec{g}|^4$ becomes increasingly subdominant compared to the trace term $\mathrm{Tr}(\mathbb{H}) / |\vec{g}|^2$ in the temperature estimator.

\subsection{Applicability of configurational temperature to HMC}

We can apply the temperature estimator to HMC simulations, provided some important clarifications are kept in mind. 
Consider the situation in which we are simulating a scalar field theory using HMC. 
The scalar field action $S[\phi]$ serves the role of a potential energy in an extended phase space. At the same time, the conjugate momenta are auxiliary variables introduced solely to facilitate efficient proposal generation.

Despite the use of a fictitious Hamiltonian dynamics, the HMC algorithm samples from the same canonical ensemble as conventional Metropolis Monte Carlo:
\begin{equation}
p(\phi) \propto e^{-S[\phi]}.
\end{equation}
The momenta are drawn from a Gaussian distribution at the beginning of each trajectory, evolve deterministically during the Hamiltonian flow, and are discarded at the end. 
Their introduction facilitates global updates and improves sampling efficiency by reducing autocorrelations compared to local-update methods, even though they do not contribute to the statistical ensemble used for observable estimation.

The artificial Hamiltonian guiding the dynamics takes the form
\begin{equation}
H[\phi, \pi] = \frac{1}{2} \sum_x \pi(x)^2 + S[\phi],
\end{equation}
where $\pi(x) \sim \mathcal{N}(0, 1)$ are auxiliary Gaussian momenta. 
After integrating out the momenta, the marginal distribution over field configurations remains canonical:
\begin{align}
p(\phi) &\propto \int \mathcal{D}\pi ~ e^{-H[\phi, \pi]} \\
&= e^{-S[\phi]} \int \mathcal{D}\pi ~ e^{- \frac{1}{2} \sum_x \pi(x)^2} \propto e^{-S[\phi]}.
\end{align}
Thus, HMC produces samples from the Boltzmann distribution $e^{-S[\phi]}$, and the configurational temperature formalism, which relies only on the distribution of configurations $\phi$, remains valid.

The temperature estimator, therefore, provides a thermodynamic consistency check for HMC simulations. 
It tests whether the field configurations $\phi$ are distributed as if drawn from a Boltzmann ensemble at the target temperature $T$. 
Since temperature is typically absorbed into the action in HMC simulations, either by working in units where $k_B T = 1$, or by introducing an explicit inverse temperature parameter $\beta$ in the action, we expect:
\begin{equation}
S[\phi] = \beta U[\phi] \quad \Rightarrow \quad T_{\rm config} \approx \frac{1}{\beta}.
\end{equation}
No information from the momenta is required to compute $T_{\rm config}$; the estimator depends only on the sampled field configurations.

\subsection{Configurational temperature estimator as a diagnostic in HMC simulations}

While the target temperature is often fixed implicitly, either by rescaling the action such that $k_B T = 1$ or by incorporating inverse temperature into the coupling constants, measuring the configurational temperature offers an independent means to verify the correctness of the simulation.

Below, we list several aspects that underscore the utility of the configurational temperature in HMC simulations.

\begin{itemize}
\item[(i.)] {\it Validation of sampling consistency}

HMC aims to generate field configurations $\phi$ according to the canonical probability distribution $p(\phi) \propto e^{-S[\phi]}$. 
However, various algorithmic issues, such as inaccurate gradient computations, incorrect acceptance criteria, integration errors, or violation of detailed balance, can lead to subtle biases in the sampled ensemble. 
Since the configurational temperature is a property of the ensemble distribution, it provides a sensitive probe for detecting such inconsistencies. 
If the simulation is thermodynamically accurate, we expect
\begin{equation}
T_{\text{config}} \approx T_{\text{input}}.
\end{equation}

\item[(ii.)] {\it Independent thermometer}

In lattice field theory, the physical temperature is often indirectly defined, for example, through the temporal lattice extent or via an effective inverse coupling. 
In such cases, verifying whether a simulation has reached equilibrium at the desired temperature is not straightforward. 
The configurational temperature provides an independent, non-kinetic estimator that can serve as a ``field-theoretic thermometer.'' 
It is advantageous in nonperturbative or strongly coupled regimes, including scalar field theories, spin systems, and lattice QCD.

\item[(iii.)] {\it Momentum-free estimator}

We can compute a {\it kinetic temperature} using the auxiliary momenta
\begin{equation}
T_{\text{kin}} = \frac{1}{N} \sum_x \left\langle \pi(x)^2 \right\rangle.
\end{equation}
However, this diagnostic merely verifies the correctness of the momentum sampling from a Gaussian distribution. 
It does not provide information about whether the field configurations are sampled correctly. 
By contrast, the configurational temperature is purely configurational and directly tests the consistency of the $\phi$-ensemble with the underlying Boltzmann distribution.

\item[(iv.)] {\it Monitoring thermalization and autocorrelations}

The configurational temperature can deviate significantly from its expected value in simulations where the system thermalizes slowly or becomes trapped in metastable states. 
Monitoring $T_{\text{config}}$ during early trajectories provides a quantitative check on equilibration. 
Persistent deviations may signal slow phase separation, poor mixing, or errors in the update dynamics or acceptance mechanism. 
Thus, the configurational temperature is a helpful indicator for diagnosing thermalization and autocorrelation issues.
\end{itemize}

In summary, the configurational temperature estimator does not replace standard validation tools such as observable comparisons or autocorrelation analyses, but it constitutes a powerful and underutilized consistency check. 
Its applicability is particularly pronounced when dealing with novel actions, large-scale simulations, or custom implementations where traditional diagnostics may be insufficient.

In the following sections, we apply the temperature estimator in compact U(1) lattice gauge theories in various dimensions.

\section{Benchmarking the Estimator in U(1) Lattice Gauge Theories}
\label{sec:u1}

We now apply the temperature estimator to the compact U(1) lattice gauge theory, a canonical Abelian model that exhibits rich phase structure in various dimensions. 
Monte Carlo studies have demonstrated that for spacetime dimensions $D < 4$, the compact U(1) lattice gauge theory exhibits only a confining phase. In contrast, for $D \geq 4$, the theory supports at least two distinct phases~\cite{Bhanot:1980pc}.

In this theory, the link variables are elements of the group U(1). 
They are associated with directed links from a site $n$ in the direction $\mu$, and are defined as
\begin{equation}
U_\mu(n) = \exp(i \theta_\mu(n)), \quad \theta_\mu(n) \in (-\pi, \pi],
\end{equation}
with $\mu = 1, \cdots, d$ denoting the spacetime directions.

The standard Wilson action for this theory takes the form
\begin{equation}
S = \beta \sum_{(\mu, \nu; n)} \left[1 - \cos\theta_{\mu\nu}(n) \right],
\end{equation}
where $\beta$ denotes the inverse coupling parameter (or temperature) and the plaquette angle $\theta_{\mu \nu}(n)$ represents the discrete curl of the link field:
\begin{equation}
\theta_{\mu \nu}(n) = \theta_\mu(n) + \theta_\nu(n + \hat{\mu}) - \theta_\mu(n + \hat{\nu}) - \theta_\nu(n),
\end{equation}
with $\hat{\mu}$ and $\hat{\nu}$ denoting the unit vectors in $\mu$ and $\nu$ directions, respectively.

The corresponding partition function is given by
\begin{equation}
Z(\beta) = \sum_{\{\theta\}} \exp\left[- S(\{\theta\})\right].
\end{equation}
The free energy per plaquette is defined as
\begin{equation}
f(\beta) = \frac{1}{N_p} \ln Z(\beta),
\end{equation}
with $N_p$ being the total number of plaquettes in the lattice.

As an observable, we consider the average plaquette energy (or action density),
\begin{equation}
E(\beta) = - \frac{\partial f(\beta)}{\partial \beta} = \left \langle 1 - \cos(\theta_{\mu\nu}) \right \rangle,
\end{equation}
which serves as an effective order parameter across different coupling regimes. 
Phase transitions are signaled by non-analytic behavior in $E(\beta)$, such as discontinuities or rapid changes. 
In practice, we track the first or second derivative, e.g., the plaquette susceptibility, as higher-order derivatives are numerically unstable. 
Throughout our study, we compute energy densities suited to each dimension and analyze them alongside the temperature estimator to understand thermal behavior across different setups.

\subsection{One-dimensional theory}

The theory uses compact link variables $U_n$, which reside on the links between lattice sites $n$ and $n + 1$.
They take values in the U(1) gauge group:
\begin{equation}
U_n = e^{i \theta_n}, \quad \theta_n \in (-\pi, \pi].
\end{equation}

The Wilson action for this system reduces to the following nearest-neighbor interaction form:
\begin{equation}
S = \beta \sum_{n = 0}^{N_\tau - 1} \left( 1 - \cos(\theta_n - \theta_{n + 1}) \right),
\end{equation}
where $N_\tau$ denotes the number of lattice sites. 

This model is equivalent to the one-dimensional classical XY model, which is exactly solvable. 
Although it does not exhibit a phase transition, it is a valuable testing ground for lattice algorithms due to its analytical tractability.

Let us denote the inverse temperature obtained through the estimator as $\hat{\beta}$.
The estimator, introduced in Eq.~\eqref{eq:another-form}, takes the form
\begin{equation}
\hat{\beta} = \frac{\sum_n h_{nn}}{\sum_n g_n^2} - \frac{2 \sum_{nm} g_n g_m h_{nm}}{\left( \sum_n g_n^2 \right)^2},
\label{eq:beta_lat}
\end{equation}
where $g_n$ and $h_{nm}$ denote the gradient and Hessian of the action, respectively:
\begin{equation}
g_n = \frac{1}{\beta} \frac{\partial S}{\partial \theta_n}, \quad h_{nm} =  \frac{1}{\beta} \frac{\partial^2 S}{\partial \theta_n \partial \theta_m}.
\label{eq:grad_hess}
\end{equation}

The explicit expressions for the gradient and Hessian terms are:
\begin{align}
g_n &= \sin(\theta_{n - 1} - \theta_n) - \sin(\theta_n - \theta_{n + 1}), \label{eq:grad} \\
h_{nn} &= \cos(\theta_{n - 1} - \theta_n) + \cos(\theta_n - \theta_{n + 1}), \label{eq:hess_diag} \\
h_{nm} &= -\cos(\theta_n - \theta_m) \delta_{m, n + 1} \nonumber \\
& ~~~ ~~~ - \cos(\theta_m - \theta_n) \delta_{m, n - 1}, \label{eq:hess_offdiag}
\end{align}
with periodic boundary conditions implied. 
These expressions allow for direct evaluation of the temperature estimator during the simulation.

Using HMC, we evolve the system in fictitious Monte Carlo time and compute observables such as the internal energy, specific heat, and the measured inverse temperature $\beta_M \equiv \langle \hat{\beta} \rangle$. 
We compute the average energy
\begin{equation}
E(\beta) = \left\langle 1 - \cos(\theta_n - \theta_{n + 1}) \right\rangle,
\label{eq:energy_lat_1d_u1}
\end{equation}
the corresponding specific heat, and the temperature estimator defined in Eq.~\eqref{eq:beta_lat}, using Eqs. \eqref{eq:grad}, \eqref{eq:hess_diag}, and \eqref{eq:hess_offdiag}. 
We compare the output $\beta_M$ obtained from the estimator with the input $\beta$ used in the simulation. We validate our results by comparing the numerically computed energy and specific heat with the exact analytical expressions.

For the theory with periodic boundary conditions, the analytical expressions for energy and specific heat are given by:
\begin{align}
E(\beta) &= 1 - \frac{I_1(\beta)}{I_0(\beta)}, \label{eq:energy_ana_1d}\\
C(\beta) &= \beta^2 \left[ 1- \frac{I_1(\beta)}{\beta I_0(\beta)} - \left( \frac{I_1(\beta)}{I_0(\beta)} \right)^2 \right],\label{eq:spec_ana_1d}
\end{align}
where $I_n(\beta)$ denotes the modified Bessel function of the first kind of order $n$. 

\begin{figure*}[t]
\centering
\includegraphics[width=0.32\textwidth]{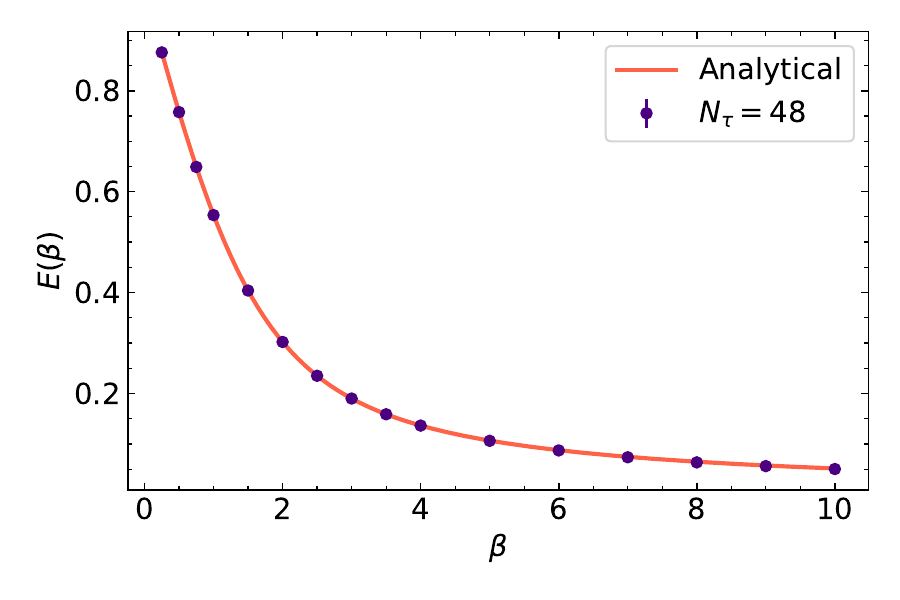}
\includegraphics[width=0.32\textwidth]{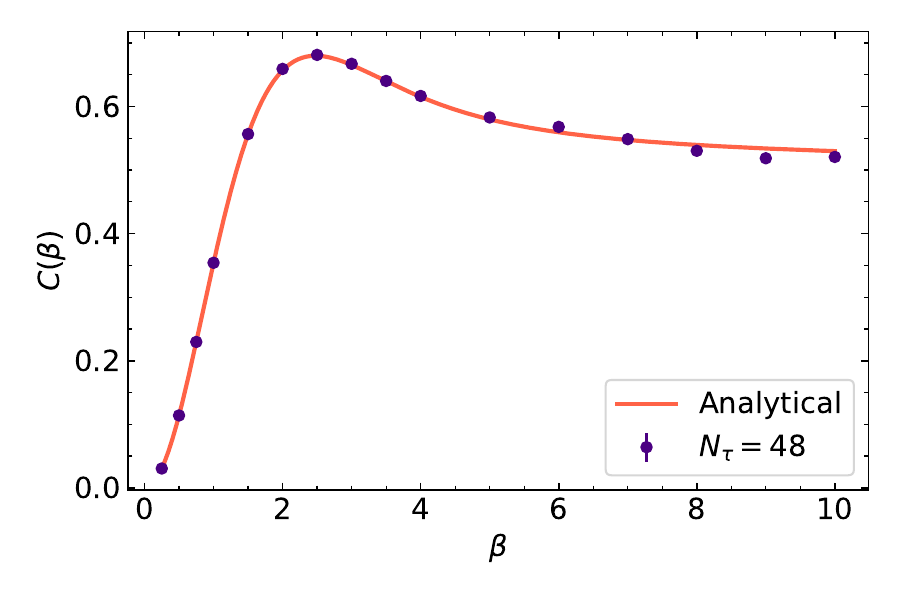}
\includegraphics[width=0.32\textwidth]{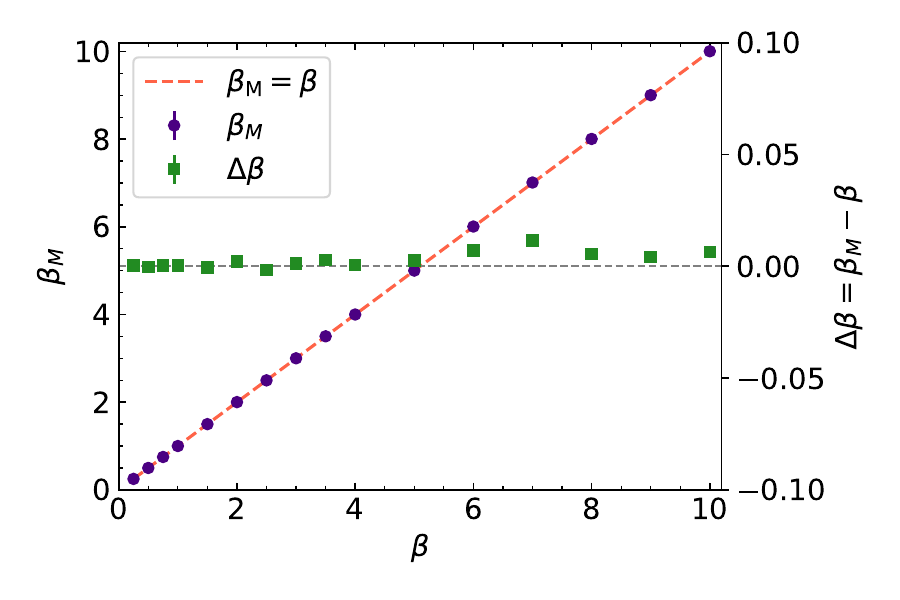}
\caption{Thermodynamic observables for the one-dimensional U(1) lattice theory on a 48 site lattice. The energy $E(\beta)$ (left), specific heat $C(\beta)$ (middle), and the measured inverse temperature $\beta_M$ (right) are plotted against the input $\beta$. 
$E(\beta)$ and $C(\beta)$ are benchmarked against the analytical results given in Eqs. ~ \eqref{eq:energy_ana_1d} and ~ \eqref{eq:spec_ana_1d}, respectively.
The plot of $\beta_M$ against $\beta$ shows the expected behavior.}
\label{fig:1d_u1_l48} 
\end{figure*}

We perform simulations on a lattice with 48 sites.
Fig.~\ref{fig:1d_u1_l48} shows the thermodynamic observables of the theory. 
The lattice energy $E(\beta)$, defined in Eq.~\eqref{eq:energy_lat_1d_u1}, and the specific heat, computed from energy fluctuations as $C(\beta) = \beta^2 N_\tau \left( \langle E^2 \rangle - \langle E \rangle^2 \right)$, are plotted as functions of the input $\beta$, and compared with the analytical results in Eqs.~\eqref{eq:energy_ana_1d} and~\eqref{eq:spec_ana_1d}. 
In addition, we show the temperature estimator $\beta_M$, obtained using Eq.~\eqref{eq:beta_lat}, and verify its agreement with the input $\beta$ across a wide range. 
To better visualize the proximity between $\beta$ and the estimated $\beta_M$, we also show the difference $\Delta\beta = \beta_M - \beta$ on a secondary axis in Fig.~\ref{fig:1d_u1_l48} (right). 
The values of $\Delta\beta$ are close to zero across the entire range, indicating consistency between input and output couplings.
The consistency of the numerical results with analytic predictions provides a strong validation of both our implementation and the effectiveness of the temperature estimator. 
The values corresponding to Fig.~\ref{fig:1d_u1_l48} are given in Table~\ref{tab:1d_u1_l48} in Appendix \ref{appendix:data}.

\begin{figure*}[ht]
\centering
\includegraphics[width=0.3\textwidth]{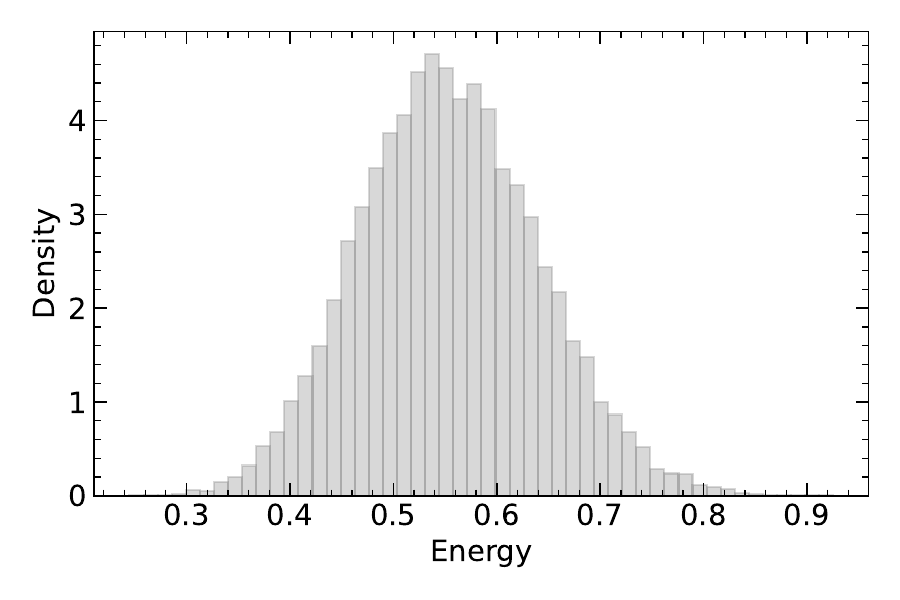}
\includegraphics[width=0.3\textwidth]{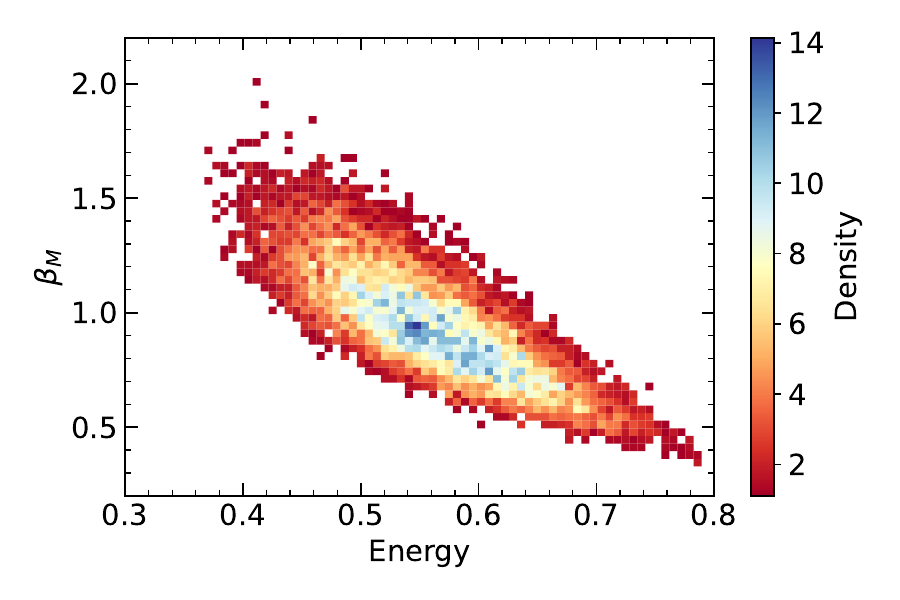}
\includegraphics[width=0.3\textwidth]{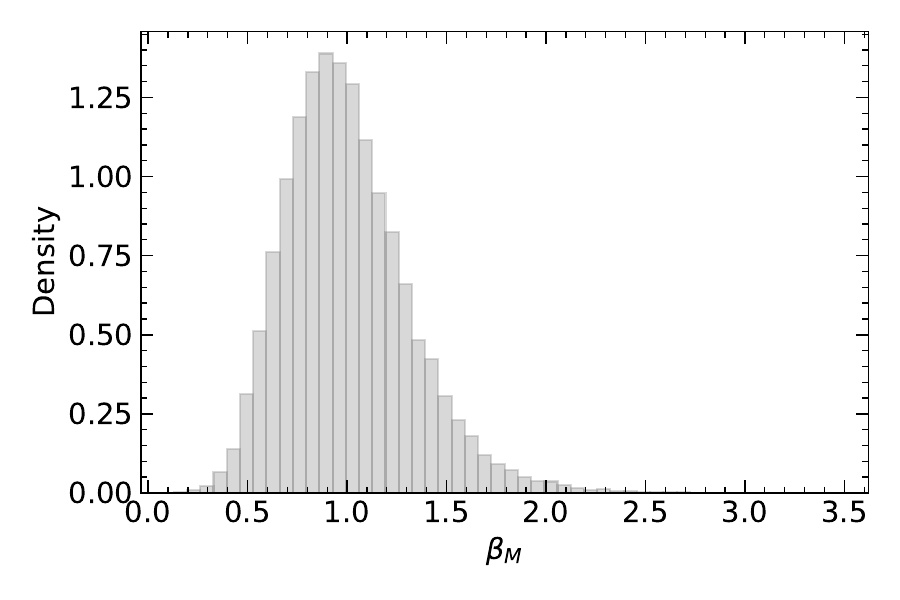}
\includegraphics[width=0.3\textwidth]{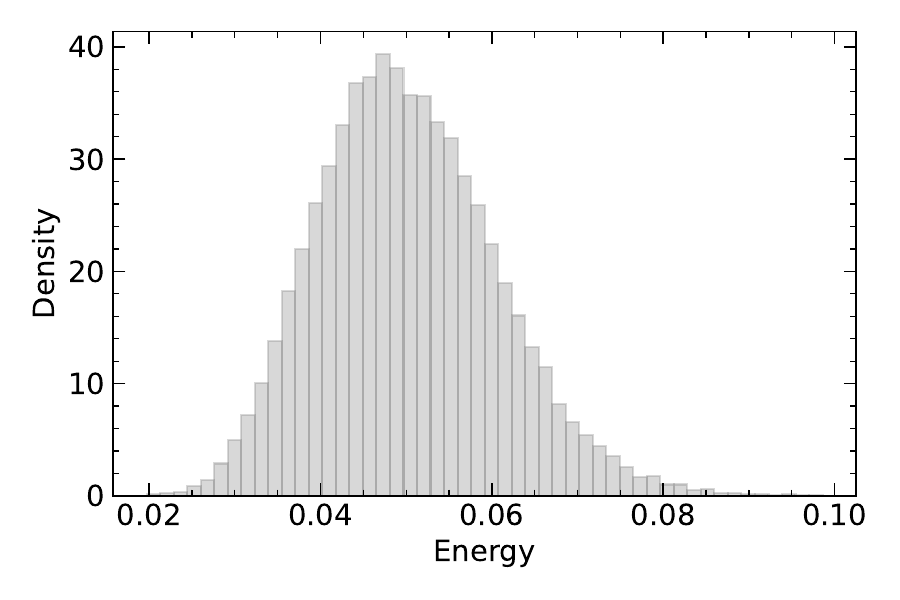}
\includegraphics[width=0.3\textwidth]{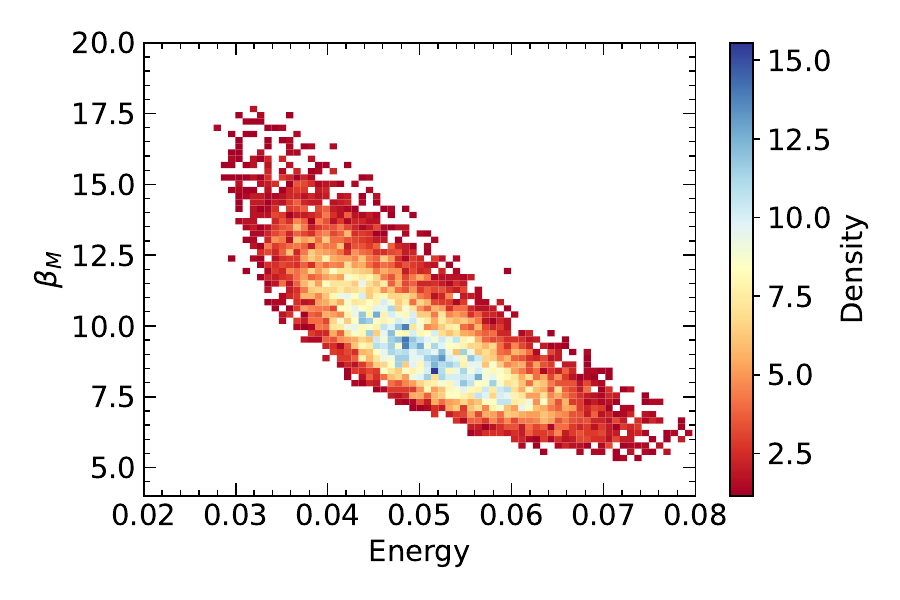}
\includegraphics[width=0.3\textwidth]{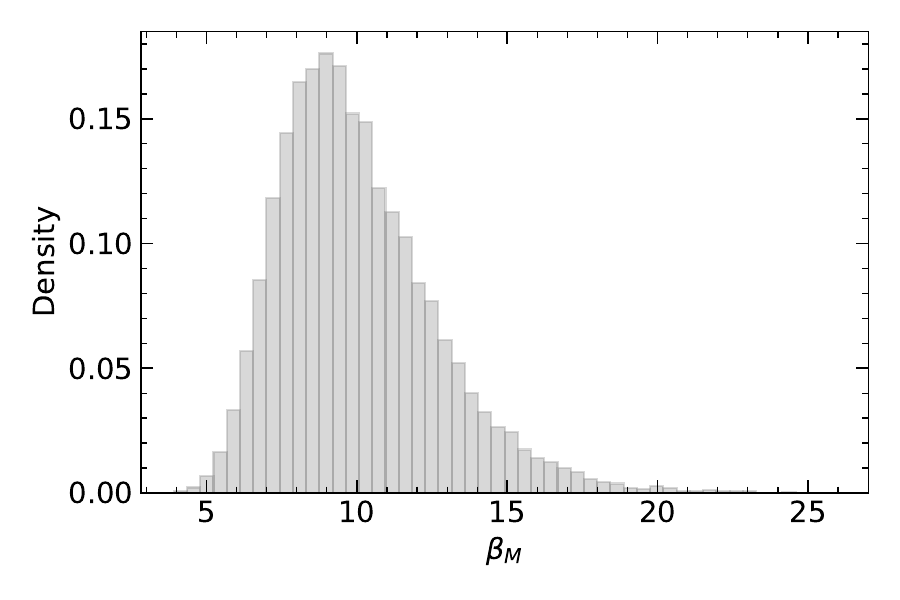}
\caption{Thermodynamic observables of the one-dimensional U(1) lattice gauge theory on a lattice with 48 sites. 
The figure presents histograms of the energy (left), the measured inverse temperature $\beta_M$ (right), and scatter plots in the form of a heat-map illustrating the relationship between energy and $\beta_M$ (middle). 
The top three plots correspond to $\beta = 10$, while the bottom three are for $\beta = 1$. 
These plots highlight the fluctuations around the average values, with the heat-map scatter plots demonstrating how energy shifts as the temperature deviates from the true value.
\label{fig:1d_u1_l48_combined}}
\end{figure*}

Figure \ref{fig:1d_u1_l48_combined} presents the thermodynamic observables for a fixed $\beta$. 
The histograms of energy and $\beta_M$ highlight fluctuations around their average values, reflecting statistical variations inherent in the Monte Carlo sampling. 
The heat-map scatter plots illustrate the relationship between energy and $\beta_M$, demonstrating how energy shifts when the temperature deviates from the true value. 

The system does not always maintain a constant temperature in numerical simulations due to statistical fluctuations. It may become trapped in local vacua or metastable states distinct from the true ground state. 
As shown in the plots, the temperature estimator method tracks these fluctuations, offering a more accurate estimate of the actual temperature during the simulation. 
By comparing the measured inverse temperature $\beta_M$ with the input $\beta$, we can assess the reliability of the simulations and ensure the system is not stuck in a false vacuum.

\begin{figure}[ht]
\includegraphics[width=0.5\textwidth]{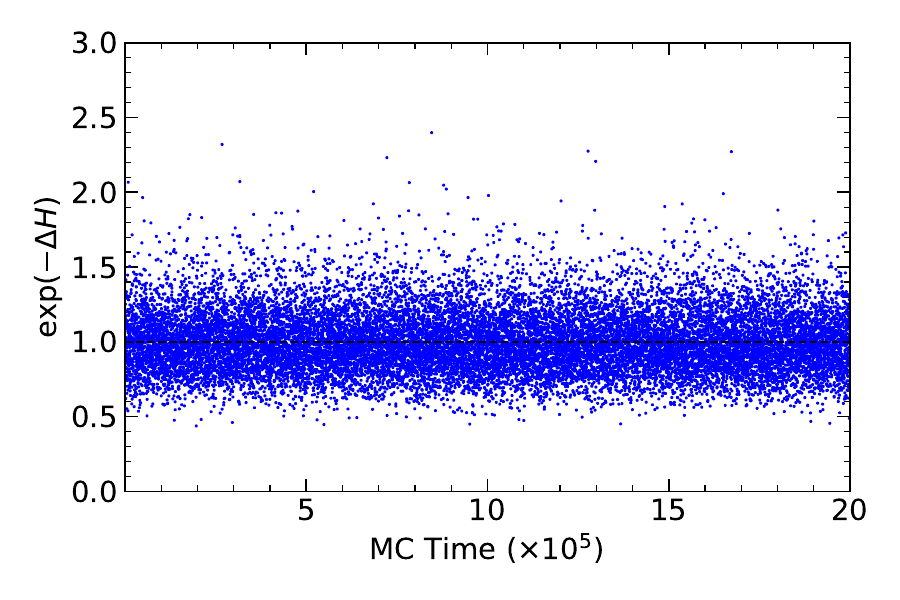}
\includegraphics[width=0.5\textwidth]{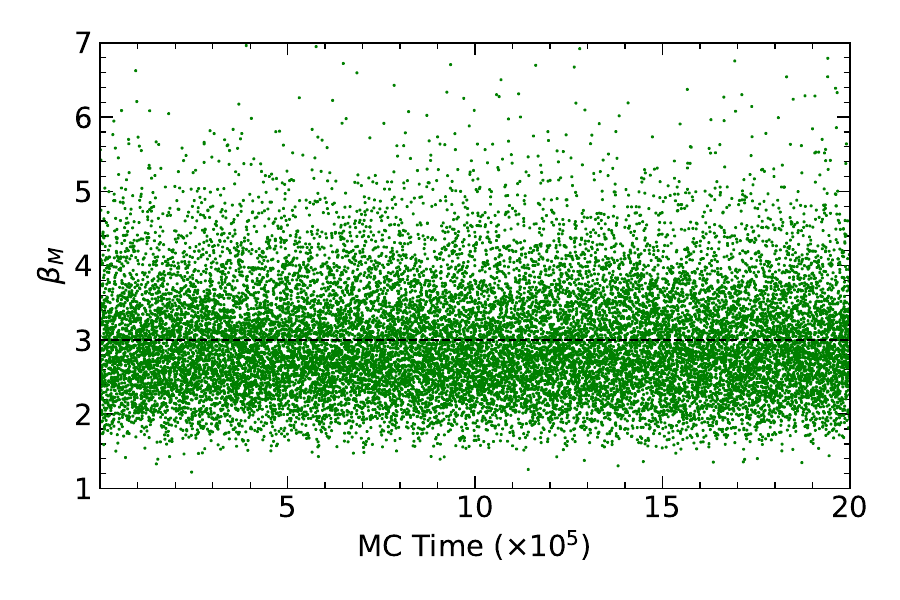}
\caption{\label{fig:1d_u1_timeseries} Time series of $\exp(-\Delta H)$ and the estimated inverse temperature $\beta_M$ for the one-dimensional U(1) lattice gauge theory on a 48-site lattice. 
The input inverse temperature is $\beta = 3$. 
}
\end{figure}

Figure~\ref{fig:1d_u1_timeseries} shows the Monte Carlo time series of the Metropolis factor $\exp(-\Delta H)$ and the estimated inverse temperature $\beta_M$ of the theory at $\beta = 3$. The data correspond to configurations recorded after thermalization. The $\exp(- \Delta H)$ values fluctuate around unity, serving as a consistency check for the Monte Carlo field updates, while $\beta_M$ tracks the estimated inverse temperature across Monte Carlo time. 

\subsection{Two-dimensional theory}

The compact U(1) lattice gauge theory in two Euclidean dimensions is defined by the Wilson action
\begin{equation}
S = \beta \sum_{(x, y)} \left[1 - \cos \theta_p(x, y)\right],
\label{eq:u1_2d_action}
\end{equation}
where $\theta_p(x, y)$ is the plaquette angle at site $(x, y)$, constructed from the oriented sum of link variables around the elementary square:
\begin{eqnarray}
\theta_p(x, y) &=& \theta_x(x, y) + \theta_y(x + 1, y) - \theta_x(x, y + 1) - \theta_y(x, y),
\end{eqnarray}
with each link variable $\theta_{x/y}(x, y) \in (-\pi, \pi]$ defined on the corresponding directed edge of the lattice. 

To extract an effective inverse temperature from field configurations, we study the local response of the action to variations in the link angles. 
The gradient of the action with respect to the $x$-oriented link $\theta_x(x, y)$ and the $y$-oriented link $\theta_y(x, y)$ is given by
\begin{align}
g_x(x, y) &= \frac{1}{\beta} \frac{\partial S}{\partial \theta_x(x, y)} \nonumber \\ &= \sin(\theta_p(x, y)) - \sin(\theta_p(x, y - 1)), \\
g_y(x, y) &= \frac{1}{\beta} \frac{\partial S}{\partial \theta_y(x, y)} \nonumber \\ &= \sin(\theta_p(x - 1, y)) - \sin(\theta_p(x, y)).
\end{align}

The corresponding diagonal components of the Hessian, which represent the second derivatives of the action with respect to the same link variable, are
\begin{align}
h_{xx}(x, y) &= \frac{1}{\beta} \frac{\partial^2 S}{\partial \theta_x(x, y)^2} \nonumber \\ &= \cos(\theta_p(x, y)) + \cos(\theta_p(x, y - 1)), \\
h_{yy}(x, y) &= \frac{1}{\beta}  \frac{\partial^2 S}{\partial \theta_y(x, y)^2} \nonumber \\ &= \cos(\theta_p(x, y)) + \cos(\theta_p(x - 1, y)).
\end{align}

Using the previously defined expressions for the gradients and diagonal Hessians with respect to the link angles, we have
\begin{align}
\hat{\beta} &= \frac{\sum_{x, y} \left[ h_{xx}(x, y) + h_{yy}(x, y) \right]}{\sum_{x, y} \left[ \left(g_x(x, y)\right)^2 + \left(g_y(x, y)\right)^2 \right]} \nonumber \\ 
&- \frac{2 \sum_{x, y, x', y'} \left[ g_x(x, y) \, g_y(x', y') \, h_{xy}(x, y; x', y') \right]}{\left( \sum_{x, y} \left[ \left(g_x(x, y)\right)^2 + \left(g_y(x, y)\right)^2 \right] \right)^2},
\label{eq:beta_lat_2d}
\end{align}
where the first term captures contributions from the diagonal elements of the Hessian, and the second term includes cross-derivative (off-diagonal) contributions. 
The notation is consistent with the site and direction-dependent link angles $\theta_x(x, y)$ and $\theta_y(x, y)$, and their corresponding local derivatives of the action. 

In the two-dimensional theory, where spatial directions are treated symmetrically, we can also independently analyze the contributions from $x$ and $y$-oriented links.  
In particular, evaluating only the first term in Eq.~\eqref{eq:beta_lat_2d} for either the $x$ or $y$-oriented links provides a reliable estimator that approaches the true coupling in the limit of large lattice volumes. 
To incorporate the contribution from off-diagonal Hessian components in the estimator of the effective coupling $\hat\beta$, we evaluate the crossed terms involving mixed second derivatives of the action. 
For each $x$-oriented link $\theta_x(x, y)$, the second derivative $\partial^2 S / \partial \theta_x(x, y) \partial \theta_y(x', y')$ is nonzero only when $\theta_y(x', y')$ shares a plaquette with $\theta_x(x, y)$. 
There are four such $\theta_y$ links associated with $\theta_x(x, y)$ through the two adjacent plaquettes: at $(x, y)$ and $(x, y - 1)$. 
The corresponding nonzero mixed second derivatives are:
\begin{align}
\frac{1}{\beta} \frac{\partial^2 S}{\partial \theta_x(x, y) \partial \theta_y(x, y)} &= - \cos(\theta_p(x, y)), \\
\frac{1}{\beta} \frac{\partial^2 S}{\partial \theta_x(x, y) \partial \theta_y(x + 1, y)} &= + \cos(\theta_p(x, y)), \\
\frac{1}{\beta} \frac{\partial^2 S}{\partial \theta_x(x, y) \partial \theta_y(x, y - 1)} &= + \cos(\theta_p(x, y - 1)), \\
\frac{1}{\beta} \frac{\partial^2 S}{\partial \theta_x(x, y) \partial \theta_y(x + 1,y - 1)} &= - \cos(\theta_p(x, y - 1)).
\end{align}
Each of these terms contributes to the total crossed term in the estimator as defined in Eq.~\eqref{eq:beta_lat_2d}. 
In it $h_{xy}(x, y; x', y')$ is nonzero only for the four neighboring $y$-oriented links $(x', y')$ listed above. 

In addition to local observables, we aim to extract non-local physical quantities, such as the string tension, by evaluating rectangular Wilson loops. 
In the exactly solvable two-dimensional U(1) lattice gauge theory, the expectation value of a Wilson loop $W(R, T)$ of spatial extent $R$ and temporal extent $T$ is given by \cite{Rothe:1987tc}
\begin{equation}
\langle W(R, T) \rangle = \left( \frac{I_1(\beta)}{I_0 (\beta)} \right)^{R T}.
\label{eq:wilson_loop_2d}
\end{equation}
This result demonstrates the area-law behavior explicitly, with the effective string tension defined as
\begin{equation}
\sigma = - \ln \left( \frac{I_1(\beta)}{I_0(\beta)} \right).
\end{equation}
In particular, taking $R = T = 1$ yields the plaquette expectation value,
\begin{equation}
P \equiv \langle \cos \theta_p \rangle = \langle W(1, 1) \rangle = \frac{I_1(\beta)}{I_0(\beta)}.
\label{eq:plaq_2d}
\end{equation}

\begin{figure*}[t]
\includegraphics[width=0.3\textwidth]{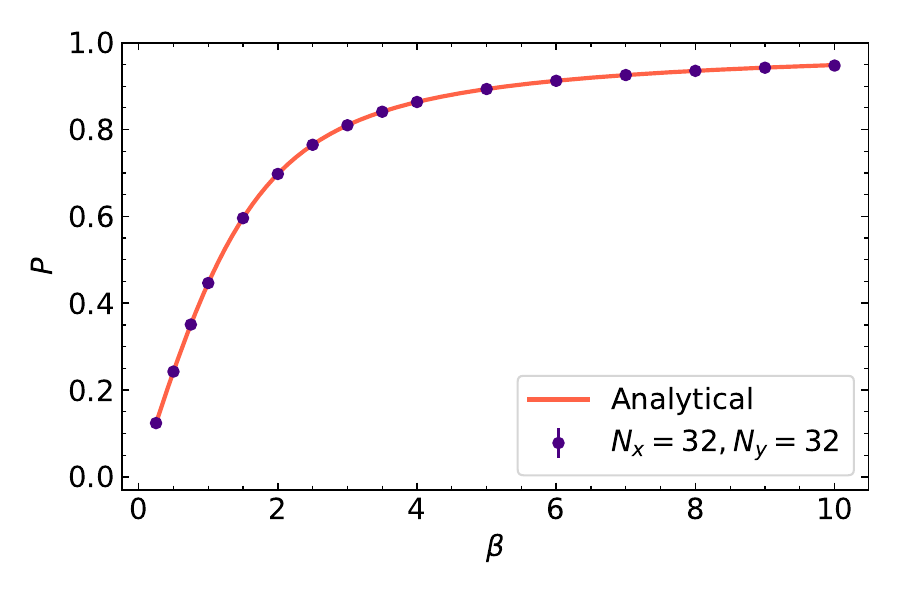}
\includegraphics[width=0.3\textwidth]{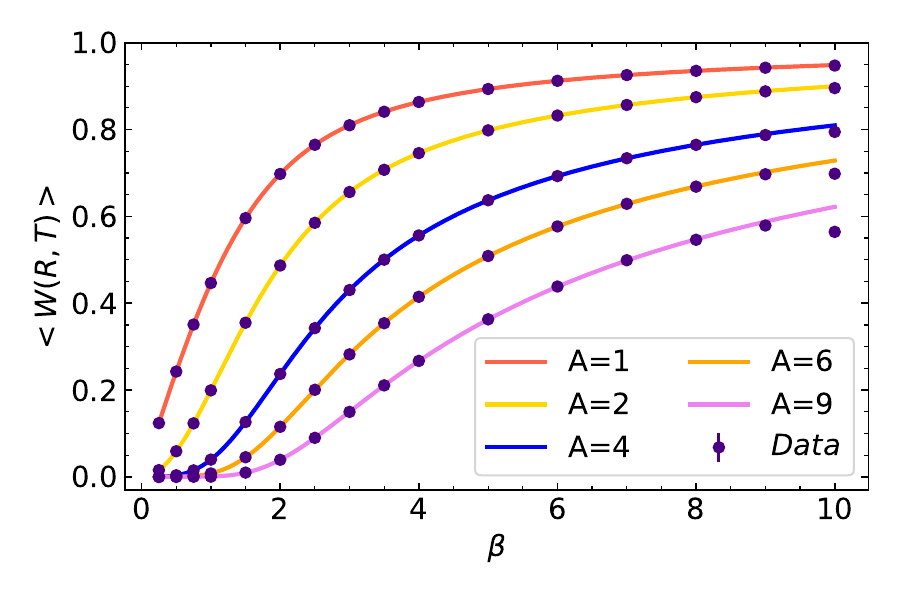}
\includegraphics[width=0.3\textwidth]{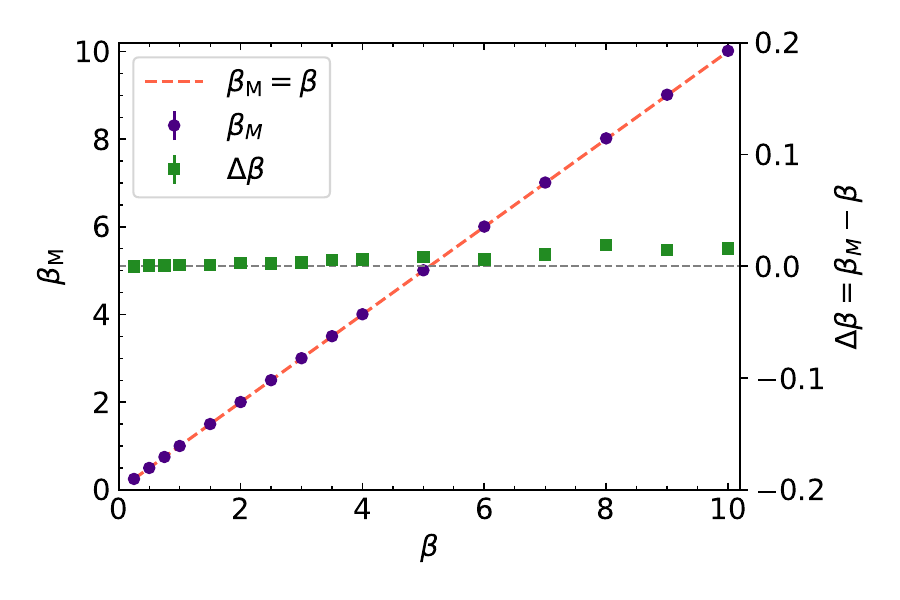}
\caption{\label{fig:plaquette_beta_2d} Two-dimensional U(1) lattice gauge theory. The expectation value of plaquette, Wilson loops of sizes $A = R \times T$ and estimated inverse temperature $\beta_M$ as functions of input $\beta$ are shown on a $32 \times 32$ lattice.
}
\end{figure*}

We show the plaquette expectation value $P$ and the estimated inverse temperature $\beta_M$ in Fig.~\ref{fig:plaquette_beta_2d} for a $32 \times 32$ lattice.  
We also detail the results in Table~\ref{tab:2d_u1_l32_l32} in Appendix \ref{appendix:data}. 
Beyond the plaquette, we can test the consistency of the field configurations by evaluating Wilson loops of larger sizes. 
Instead of only the $1 \times 1$ plaquette, we compute expectation values of rectangular Wilson loops $W(R, T)$ with various $R$ and $T$ on the same $32 \times 32$ lattice. 
These are then compared with the corresponding analytical predictions as given in Eq. \eqref{eq:wilson_loop_2d} in Fig.~\ref{fig:plaquette_beta_2d}. 
While the analytical expression for the Wilson loop is valid for all values of $\beta$, in practice, numerical deviations can appear at large $\beta$, especially for Wilson loops with large area $R T$. 
This behavior is evident in Fig.~\ref{fig:plaquette_beta_2d}. 
For a large area $R T$, even a slight deviation from unity in the base gets exponentially amplified in the exponent. 
For instance, if we write $\left(1 - \epsilon\right)^{R T} \approx e^{-\epsilon R T}$, then even a tiny $\epsilon > 0$ leads to significant suppression when $R T$ is large. 
It explains why Wilson loop values can become extremely small and prone to numerical underestimation in practice, particularly at large $\beta$ and for large loop areas. 

The results, shown in Fig.~\ref{fig:plaquette_beta_2d}, demonstrate that the estimated inverse temperature from the observable is consistent with the input value, confirming the reliability of the temperature estimator.
Although scatter plots between output observables and the estimated inverse temperature $\beta_M$ are not shown here for the two-dimensional case, as we did in the one-dimensional case in Fig.~\ref{fig:1d_u1_l48_combined}, such plots are helpful to visualize how fluctuations and metastable states affect the system’s effective temperature during simulations. 
The temperature estimator $\beta_M$ remains an important tool to monitor these effects and assess the reliability of the results.

\begin{figure}[ht]
\includegraphics[width=0.5\textwidth]{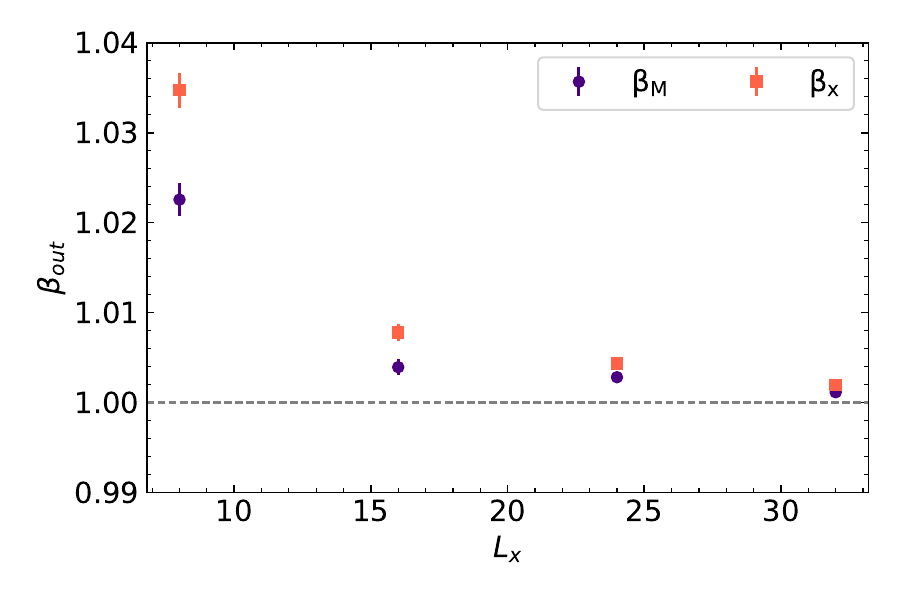}
\includegraphics[width=0.5\textwidth]{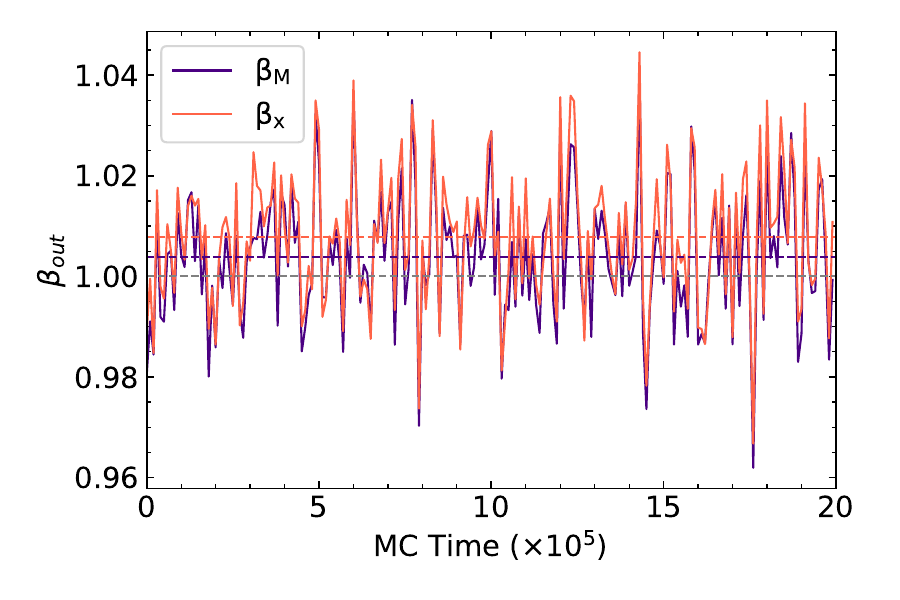}
\caption{\label{fig:beta_vol_2d} (Left) Comparison of the full estimator $\beta_M$ and the directional estimator $\beta_x$ as a function of lattice size $L_x$ for $L_x \times L_x$ lattices at $\beta = 1$. 
(Right) The Monte Carlo time history of both estimators on a $16 \times 16$ lattice illustrates fluctuations and near agreement between the two estimators over simulation time.
The dotted lines correspond to the average values of the respective observables over the Monte Carlo history.}
\end{figure}

To analyze the dominant contributions to the extracted inverse temperature, we consider a simplified estimator 
\begin{equation}
\beta_x = \left \langle \frac{\sum_{x, y} h_{xx}(x, y)}{\sum_{x, y} \left(g_x(x, y)\right)^2} \right \rangle,
\end{equation}
constructed solely from the diagonal terms involving derivatives with respect to $x$-oriented links. 
By comparing this with the full estimator $\beta_M$, which includes diagonal and off-diagonal Hessian contributions, we analyze the relative importance of the off-diagonal terms. 
At larger lattice volumes, the diagonal terms dominate, and $\beta_x$ approximates the input $\beta$. 
However, the off-diagonal terms become more significant for smaller volumes and must be included for a more accurate estimate. 
In Fig.~\ref{fig:beta_vol_2d}, we show the behavior of the full estimator $\beta_M$ and the simplified directional estimator $\beta_x$ for $\beta = 1$ across different lattice sizes. 
This comparison illustrates that the off-diagonal terms contribute very little to the estimator for larger volumes, and $\beta_M$ closely approaches the input coupling. 
In contrast, even the full estimator $\beta_M$ for small volumes deviates from the expected value, indicating that the temperature estimator becomes more reliable in the large-volume limit.

Hence, for larger lattice volumes, one can reliably use the simpler directional estimator as a proxy for the inverse temperature, since it provides an accurate approximation in the large-volume limit where the off-diagonal contributions become negligible. 
In the 4D case considered next, which effectively corresponds to a larger hypervolume than 2D, we omit the off-diagonal terms in the temperature estimator to reduce computational complexity and because their contribution is expected to be negligible in the larger 4D space.

\subsection{Four-dimensional theory}

The four-dimensional compact U(1) lattice gauge theory is one of the simplest gauge theories exhibiting a nontrivial phase transition. 
Despite its apparent simplicity, the theory features a rich phase structure, transitioning from a confining phase at strong coupling to a Coulomb-like phase at weak coupling. 
This makes it a valuable testbed for exploring finite-temperature behavior and gauging the performance of our temperature estimator in a controlled but nontrivial setting.

The compact U(1) lattice gauge theory in four Euclidean dimensions is defined by the Wilson action
\begin{equation}
S = \beta \sum_x \sum_{\mu < \nu} \left[ 1 - \cos\theta_{\mu\nu}(x) \right],
\label{eq:u1_4d_action}
\end{equation}
where $x$ labels the lattice sites and the indices $\mu, \nu \in \{0,1,2, 3\}$ denote the directions in Euclidean spacetime. 
The plaquette angle $\theta_{\mu\nu}(x)$ is constructed as the oriented sum of link variables around the elementary square in the $\mu$-$\nu$ plane:
\begin{equation}
\theta_{\mu \nu}(x) = \theta_\mu(x) + \theta_\nu(x + \hat\mu) - \theta_\mu(x + \hat\nu) - \theta_\nu(x),
\end{equation}
with $\theta_\mu(x) \in (-\pi, \pi]$ being the compact U(1) link variable along the $\mu$-direction at site $x$.

There are $\binom{4}{2} = 6$ such plaquettes per site, corresponding to the planes: $01$, $02$, $03$, $12$, $13$, and $23$. 
The sum over $\mu < \nu$ ensures each plaquette is counted once.
Each link contributes to six plaquettes, and the local structure of the action allows us to compute its derivatives with respect to individual link angles to obtain local observables such as the gradient and the Hessian.

For this theory, there is a phase transition around $\beta \approx 1.01$, which has been shown through Monte Carlo simulations in the literature \cite{Arnold:2000hf, Bonati:2013ota, Bhanot:1981zg, Lautrup:1980xr}. 
In this work, our primary objective is not to obtain a high-precision determination of the critical coupling. 
Instead, we aim to evaluate the effectiveness of our proposed temperature estimator in capturing the thermal behavior of the system across a broad range of couplings. 
By applying the estimator to gauge field configurations generated over a wide range of $\beta$, we analyze its robustness and physical sensitivity, particularly its ability to signal transition-like behavior, without relying on traditional order parameters.

We first investigate the phase structure of this theory by tuning the coupling $\beta$ and studying various gauge-invariant observables that characterize the confinement-deconfinement transition. 
The average plaquette,
\begin{equation}
\langle P \rangle = \frac{1}{6V} \sum_x \sum_{\mu < \nu} \cos\theta_{\mu\nu}(x),
\end{equation}
serves as a basic diagnostic of gauge field fluctuations and exhibits a sharp increase near the critical coupling. 
We also monitor its susceptibility
\begin{equation}
\chi_P = V \left( \langle P^2 \rangle - \langle P \rangle^2 \right),
\end{equation}
which enhances sensitivity to critical behavior.

In addition to these conventional observables, we define a local, scale-invariant estimator for the inverse coupling:
\begin{equation}
\hat{\beta} = \frac{\sum_x \sum_\mu h_{\mu \mu}(x)}{\sum_x \sum_\mu \left[ g_\mu(x) \right]^2},
\end{equation}
where $g_\mu(x)$ and $h_{\mu\mu}(x)$ denote the local gradient and diagonal Hessian of the action with respect to the link variable $\theta_\mu(x)$. 
As discussed in the two-dimensional case, we neglect the contributions from the off-diagonal entries in the Hessian when constructing the estimator.

\begin{figure*}[t]
\centering
\includegraphics[width=0.32\textwidth]{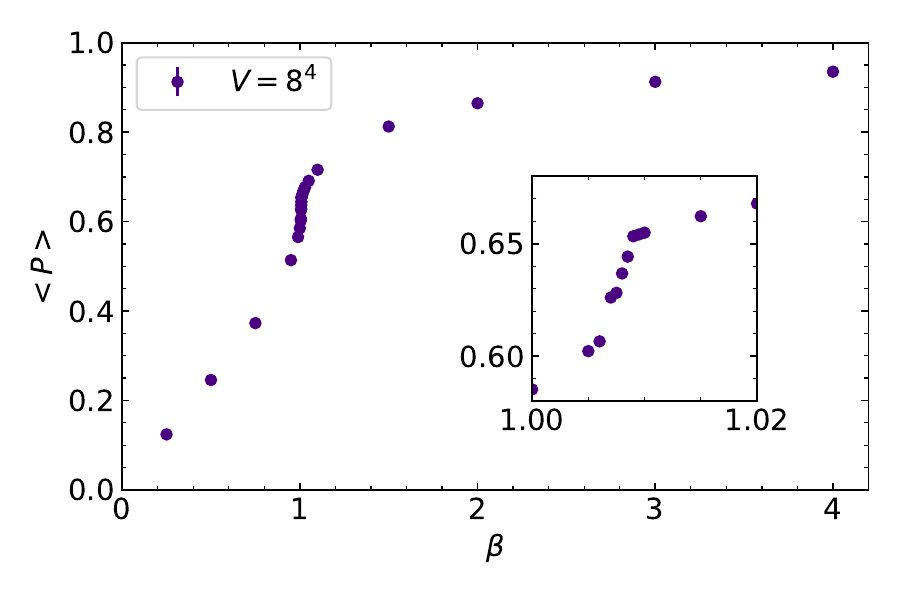}
\includegraphics[width=0.32\textwidth]{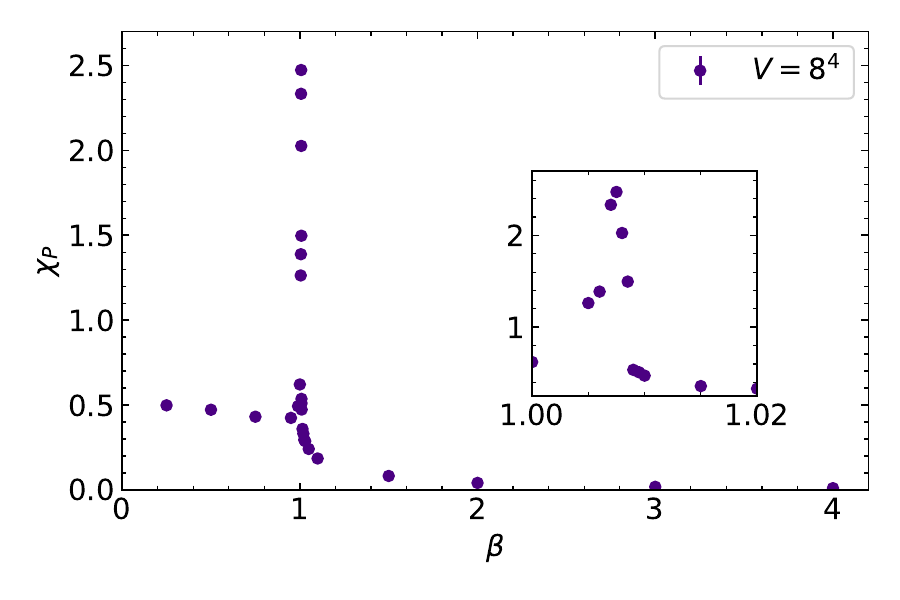}
\includegraphics[width=0.32\textwidth]{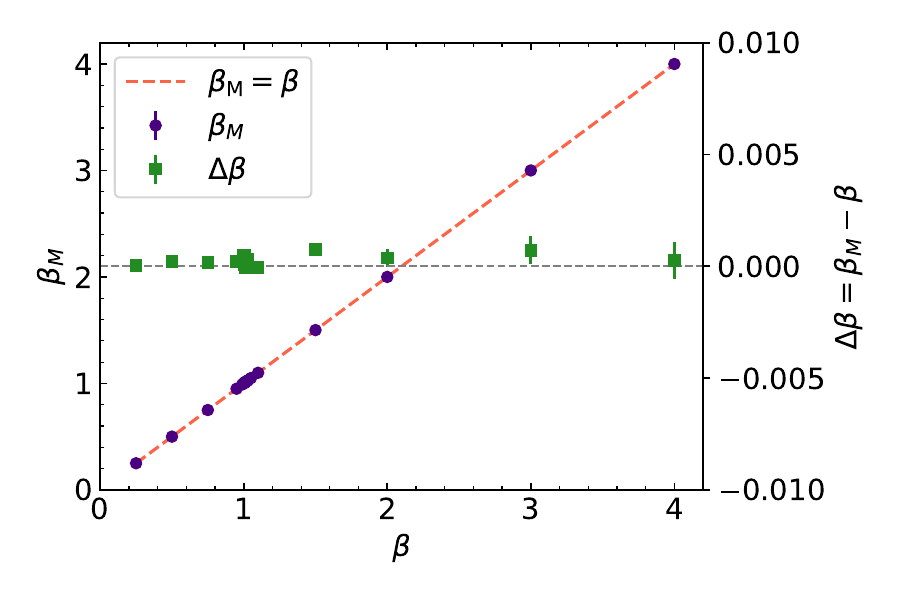}
\caption{Thermodynamic observables for the four-dimensional U(1) lattice theory on an $8^4$ lattice. The plaquette $\langle P \rangle$ (left), the plaquette susceptibility $\chi_P$ (middle), and the estimated inverse temperature $\beta_M$ (right) are plotted against the input temperature $\beta$.}
\label{fig:4d_u1_8} 
\end{figure*}

Fig. \ref{fig:4d_u1_8} presents the thermodynamic observables of the theory on an $8^4$ lattice. 
We show the plots of plaquette $\langle P \rangle$ (left), the plaquette susceptibility $\chi_P$ (middle), and the estimated inverse temperature $\beta_M$ against the input temperature $\beta$ (right). 
For this lattice size, our simulations locate the transition near $\beta \approx 1.0075$. 
Remarkably, even without incorporating the off-diagonal contributions of the Hessian, the temperature estimator reproduces the input $\beta$ with high accuracy. 
This agreement reinforces the correctness of the simulations and supports the robustness of the estimator in capturing the thermal scale. 
The values corresponding to Fig.~\ref{fig:4d_u1_8} are given in Table~\ref{tab:4d_u1_l8} in Appendix \ref{appendix:data}. 
In Fig.~\ref{fig:4d_u1_combined} we show the histograms of the plaquette and $\beta_M$ that exhibit fluctuations around their mean values, reflecting statistical variations from Monte Carlo sampling. 
The central heat-map illustrates the correlation between plaquette and $\beta_M$, capturing how energy responds to changes in the temperature.

\begin{figure*}[ht]
\centering
\includegraphics[width=0.3\textwidth]{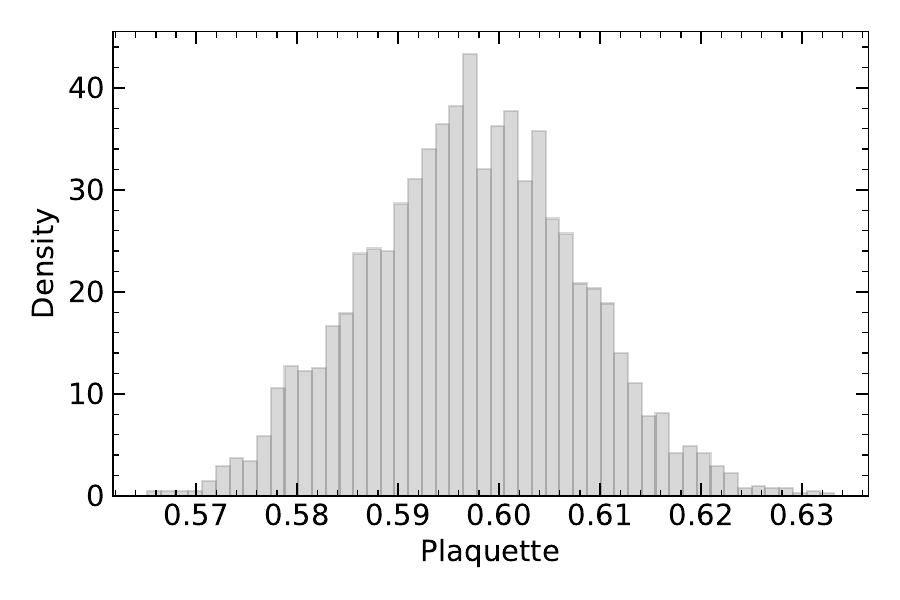}
\includegraphics[width=0.3\textwidth]{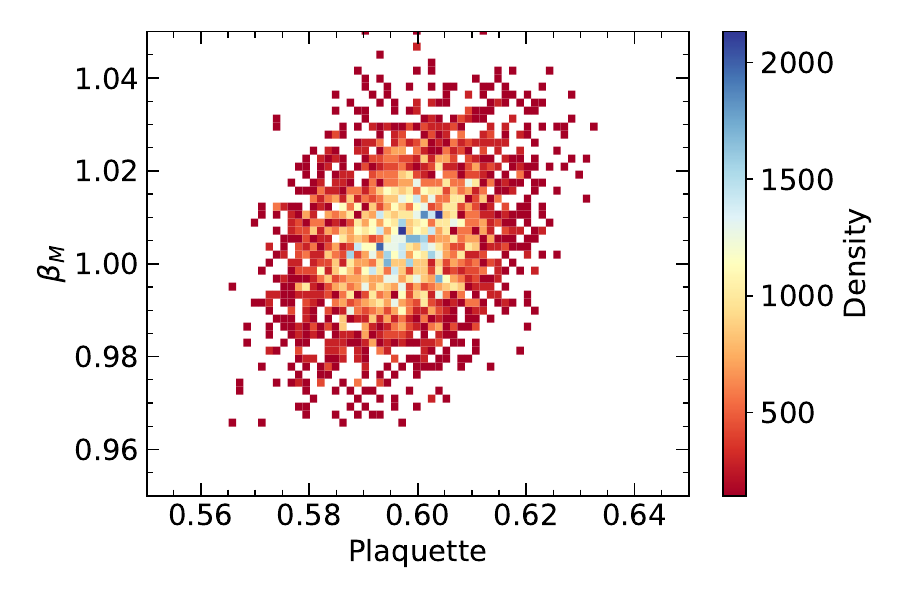}
\includegraphics[width=0.3\textwidth]{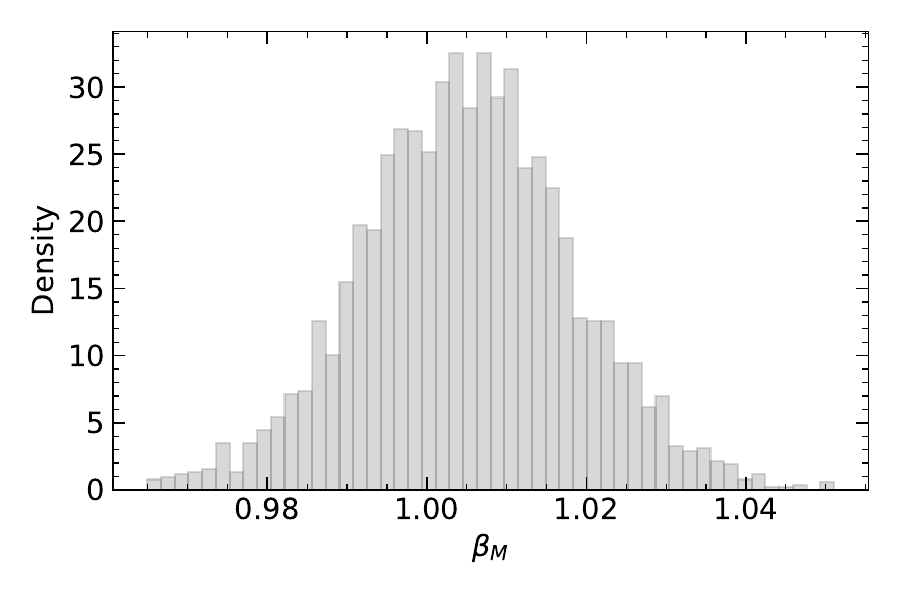}
\includegraphics[width=0.3\textwidth]{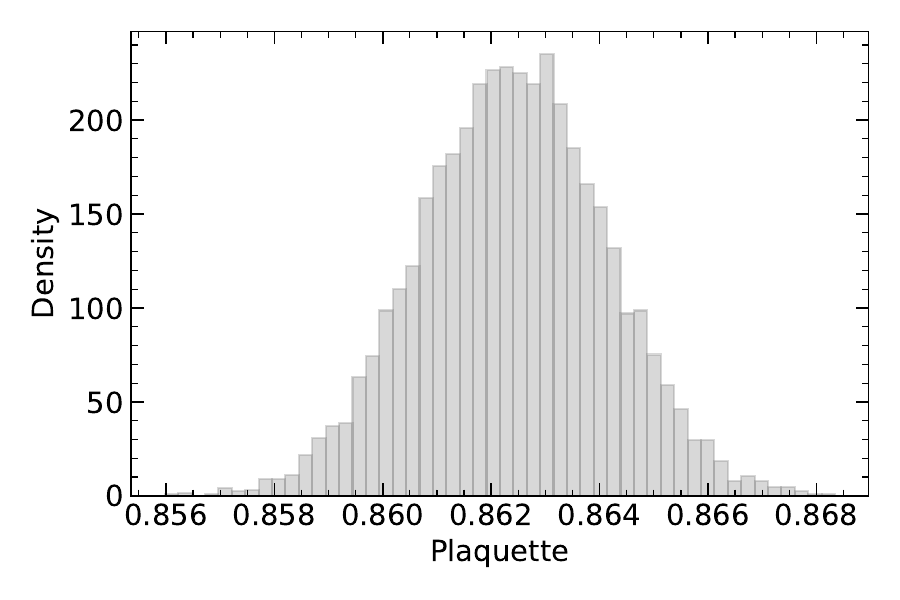}
\includegraphics[width=0.3\textwidth]{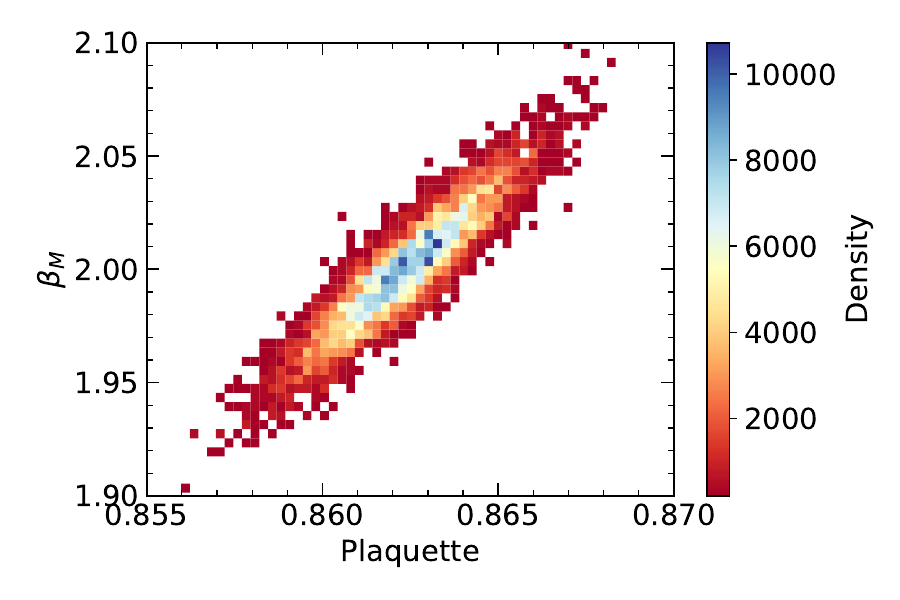}
\includegraphics[width=0.3\textwidth]{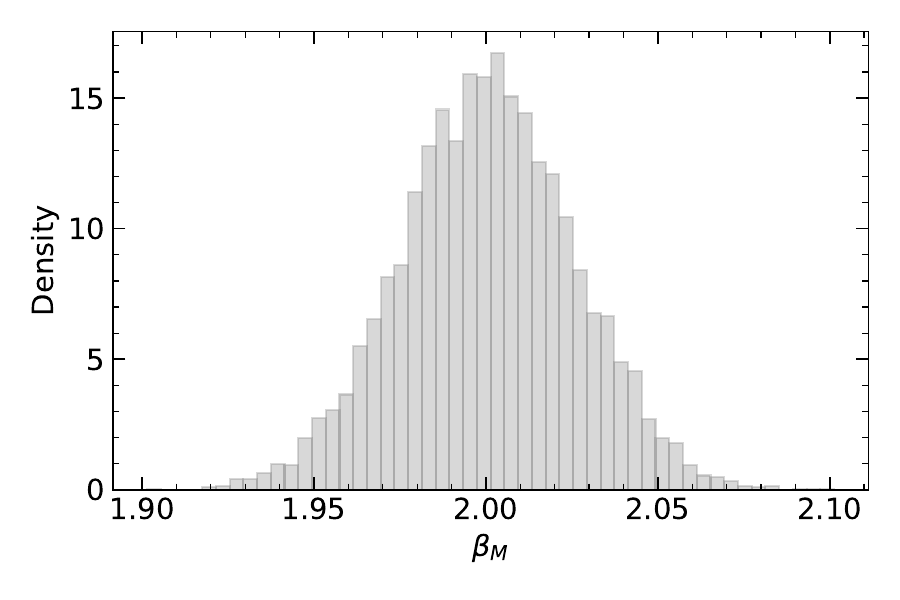}
\caption{Thermodynamic observables for the four-dimensional U(1) lattice theory with $V = 8^4$. 
The figure presents histograms of the plaquette (left), the measured inverse temperature $\beta_M$ (right), and a scatter plot in the form of a heat-map illustrating the relationship between plaquette and $\beta_M$ (middle). 
The top three plots correspond to $\beta = 1.005$, the critical $\beta$, while the bottom three are for $\beta = 2.0$. 
\label{fig:4d_u1_combined}}
\end{figure*}

We present the numerical observations from this four-dimensional case study, highlighting how the estimator reflects simulation behavior.

\begin{itemize}
\item \textit{Thermalization:} 
The estimator can be a valuable indicator for checking simulation thermalization. 
In the early stages of a Monte Carlo run, observables such as the plaquette and $\beta_M$ exhibit similar thermalization behavior, gradually settling into stable distributions, as seen in Fig.~\ref{fig:therm}. 
Even with early statistics, the estimator $\beta_M$ tracks this transition reliably, making it a useful diagnostic tool for probing thermalization in the simulation.

\begin{figure}[ht]
\centering
\includegraphics[width=0.7\textwidth]{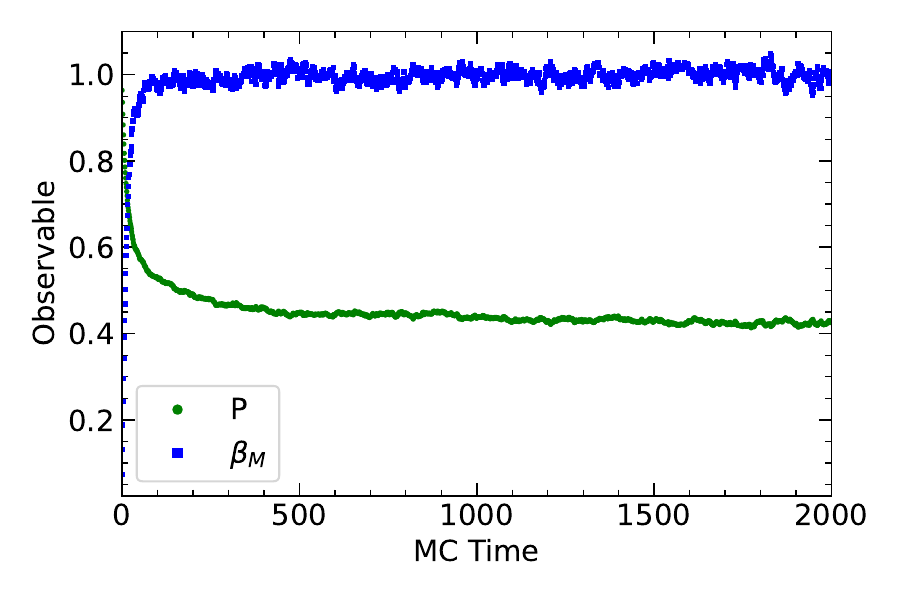}
\caption{The initial part of Monte Carlo time history, including thermalization, of the plaquette and $\beta_M$ on an $8^4$ lattice at $\beta = 1.0$.
\label{fig:therm}}
\end{figure}

\item \textit{Error diagnosis:} 
The estimator can also be used as a diagnostic tool to identify implementation errors in the simulation. 
As a demonstration, we intentionally introduce a flaw in the Metropolis accept-reject step by drawing the random number from a uniform distribution in the range $[0.5, 1.5]$ instead of the correct interval $[0, 1]$. 
This modification alters the acceptance probability, leading to systematic deviations in the simulation outcome. 
In particular, the extracted value of $\beta_M$ from the estimator no longer matches the input value used in the simulation. 
This discrepancy is visible in Fig.~\ref{fig:check}, where we compare the estimator results from the correct and incorrect implementations.

\begin{figure}[ht]
\centering
\includegraphics[width=0.7\textwidth]{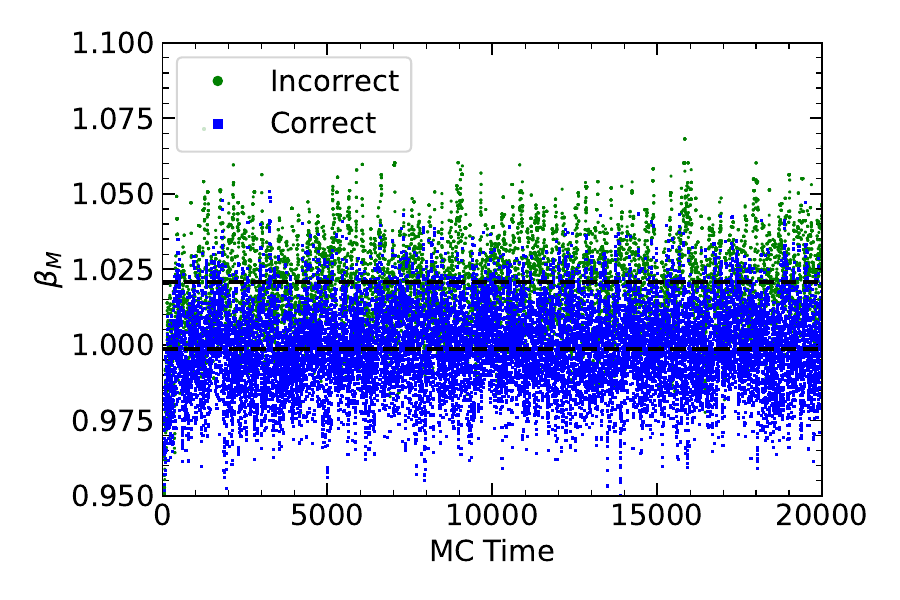}
\caption{Comparison of the extracted $\beta_M$ from simulations using the correct and incorrect Metropolis updates, performed at input $\beta = 1.0$. The estimator captures the deviation in the incorrect case, demonstrating its sensitivity to errors in the algorithm.
\label{fig:check}}
\end{figure}

\item \textit{Phase transition order parameter?:} 
While the estimator $\beta_M$ serves well as a diagnostic tool for thermalization and error detection, it is unsuitable as an order parameter for detecting phase transitions. 
In Fig.~\ref{fig:order}, we show the Monte Carlo time histories of the plaquette and $\beta_M$ at the critical coupling $\beta = 1.0075$. 
The plaquette exhibits clear signs of phase coexistence, indicating the presence of a first-order transition, whereas $\beta_M$ remains smooth and does not indicate distinct phases. 
This demonstrates that $\beta_M$ is insensitive to phase separation and is therefore not appropriate as an order parameter for studying critical behavior.

\begin{figure}[ht]
\centering
\includegraphics[width=0.7\textwidth]{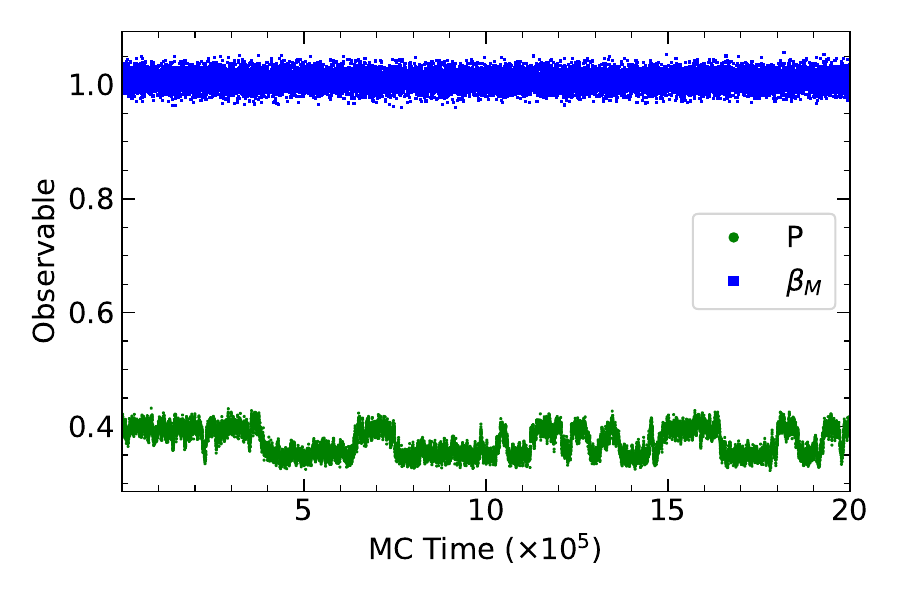}
\caption{Monte Carlo time histories of the plaquette and $\beta_M$ at the critical coupling $\beta = 1.0075$ on an $8^4$ lattice. 
\label{fig:order}}
\end{figure}

\end{itemize}

\section{Conclusions and Future Directions}
\label{sec:conc}

In this work, we build on Rugh’s geometric formulation of temperature in the microcanonical ensemble \cite{PhysRevLett.78.772}, which expresses temperature in terms of phase-space geometry. 
We systematically assess a temperature estimator that incorporates both gradient and Hessian information within the context of lattice gauge theories. 
By deriving a local, configuration-only expression rooted in Rugh’s entropy-based framework, we construct a gauge-invariant diagnostic that is independent of momenta and well suited for use in equilibrium Monte Carlo simulations.

Our investigations across 1D, 2D, and 4D compact U(1) lattice gauge theories demonstrate that the estimator reliably recovers the input inverse temperature across various couplings and lattice sizes. 
The configurational temperature closely matched theoretical predictions in the analytically tractable 1D and 2D models, validating its stability and sensitivity. 
In the 4D setting, we have a system that undergoes a phase transition. 
Here, the estimator also captures the input temperature for coupling values across the transition.

The configurational temperature estimator provides a complementary diagnostic that can reveal subtle violations of equilibrium, such as metastability or poor sampling. 
While standard observables like the average plaquette will eventually reflect the correct temperature upon complete equilibration, they may appear deceptively stable in partially thermalized or poorly sampled ensembles. 
In such cases, the configurational temperature can act as an early indicator of underlying inconsistencies.
Moreover, we demonstrated that for large lattices, the off-diagonal Hessian contributions to the estimator become negligible, enabling simplified implementations without significant loss of accuracy.

Building on the success of the current study, we note that several promising avenues can emerge: $(i.)$ {\it Extension to non-Abelian gauge theories:} Applying the estimator to SU($N$) lattice gauge theories, particularly in QCD-like settings, would test its robustness in more complex, non-Abelian configurations and offer new tools for diagnosing confinement and chiral symmetry breaking; $(ii.)$ {\it Anisotropic and finite-temperature lattices:} The estimator can be leveraged to independently verify temperature anisotropies, especially in systems with nonuniform temporal and spatial lattice spacings, or for simulations at finite temperature and chemical potential; $(iii.)$ {\it Integration into HMC and ML-accelerated algorithms:} Embedding the estimator into HMC or machine learning-guided samplers could offer real-time feedback on sampling accuracy, potentially improving convergence and reducing computational overhead; and $(iv.)$ {\it Investigation of critical slowing down and topological freezing:} By tracking temperature fluctuations, the estimator could offer new insights into autocorrelation times near criticality or topological sectors that are notoriously hard to sample.

In finite-temperature QCD, the inverse temperature is controlled by the temporal extent of the lattice: $T = 1/(N_\tau a)$, where  $N_\tau$ is the number of lattice sites in the temporal direction, and $a$ is the lattice spacing.
Ensuring that simulations indeed sample from the correct thermal distribution is critical, especially near phase transitions such as the deconfinement crossover in QCD.
However, in QCD, the temperature is often indirectly inferred via the input coupling (e.g., $\beta$), which is subject to renormalization and scale-setting ambiguities.
The configurational temperature estimator provides a way to independently measure the effective temperature from field configurations, offering a non-perturbative thermometer that could help verify thermalization in QCD simulations, check for systematic errors, or identify exceptional configurations that may signal systematic issues in large-scale finite-temperature QCD runs, and refine temperature determination near the critical region (chiral/deconfinement transitions).

The QGP studied at RHIC and LHC exists in a regime of extreme temperature and strong coupling, where lattice QCD is the primary non-perturbative tool.
Observables like Polyakov loops, susceptibilities, and quark condensates are sensitive to thermal fluctuations. 
The temperature estimator could serve as a confidence check for thermal initialization, detect slow thermalization or metastability in ensemble generation, and help compare the effective temperature derived from the lattice against the input $T$, crucial for hydrodynamic model matching.

\acknowledgments
The work of A.J. was supported in part by a Start-up Research Grant from the University of the Witwatersrand. 
A.J. gratefully acknowledges the warm hospitality of the National Institute for Theoretical and Computational Sciences (NITheCS) and Stellenbosch University during the NITheCS Focus Area Workshop, {\it Decoding the Universe: Quantum Gravity and Quantum Fields.} 
N.S.D. gratefully acknowledges The Institute of Mathematical Sciences (IMSc), Chennai, for providing support and resources during the course of this study. 
V.L. acknowledges the support and computational facilities provided by the Indian Institute of Science Education and Research - Mohali.

\appendix

\section{Data Tables for 1D, 2D, and 4D Compact U(1) Lattice Gauge Theories}
\label{appendix:data}

\begingroup
\renewcommand*{\arraystretch}{1.5}
\begin{table}[ht]
\centering
\begin{tabular}{|p{0.075\textwidth}|p{0.15\textwidth}|p{0.15\textwidth}|p{0.16\textwidth}|}\hline \hline 
~~$\beta$~~ & ~~$E$~~     & ~~$C_v$~~ &~~$\beta_M$~~\\ \hline \hline   
10.00  &  0.05030 (1)  & 0.52069 (354) & 10.00640 (185)\\
 9.00  &  0.05606 (1)  & 0.51860 (353) & 9.00409 (165)\\
 8.00  &  0.06346 (1)  & 0.53036 (357) & 8.00558 (147)\\
 7.00  &  0.07355 (1)  & 0.54866 (363) & 7.01159 (129)\\
 6.00  &  0.08718 (1)  & 0.56794 (369) & 6.00708 (110)\\
 5.00  &  0.10621 (2)  & 0.58282 (374) & 5.00257 (91)\\
 4.00  &  0.13640 (2)  & 0.61667 (385) & 4.00048 (72)\\
 3.50  &  0.15879 (2)  & 0.64027 (392) & 3.50280 (63)\\
 3.00  &  0.18995 (3)  & 0.66721 (400) & 3.00122 (54)\\
 2.50  &  0.23509 (3)  & 0.68122 (404) & 2.49835 (45)\\
 2.00  &  0.30207 (4)  & 0.65918 (398) & 2.00211 (37)\\
 1.50  &  0.40394 (5)  & 0.55666 (366) & 1.49952 (29)\\
 1.00  &  0.55343 (6)  & 0.35419 (292) & 1.00035 (22)\\
 0.75  &  0.64889 (7)  & 0.22962 (235) & 0.75046 (19)\\
 0.50  &  0.75751 (7)  & 0.11398 (165) & 0.49976 (17)\\
 0.25  &  0.87581 (7)  & 0.03050 (86) & 0.25029 (15)\\
\hline \hline
\end{tabular}
\caption{\label{tab:1d_u1_l48}{One-dimensional compact U(1) lattice gauge theory on a 48-site lattice. Summary of the thermodynamic observables: inverse temperature $\beta$, energy $E$, specific heat $C_v$, and estimated inverse temperature $\beta_M$.}}
\end{table} 
\endgroup

\begingroup
\renewcommand*{\arraystretch}{1.5}
\begin{table}[ht]
\centering
\begin{tabular}{|p{0.075\textwidth}|p{0.15\textwidth}|p{0.16\textwidth}|} \hline \hline 
~~$\beta$~~ & ~~$P$~~   &~~$\beta_M$~~\\ \hline \hline   
    10.00 & 0.94743 (2)& 10.01600 (342) \\
    9.00  &0.94249 (2)& 9.01442 (309) \\
    8.00  &0.93525 (2)& 8.01902 (272) \\
    7.00  &0.92554 (2)& 7.01052 (237) \\
    6.00  &0.91231 (3)& 6.00628 (202) \\
    5.00  &0.89340 (3)& 5.00843 (169) \\
    4.00  &0.86352 (4)& 4.00625 (136) \\
    3.50  &0.84108 (5)& 3.50550 (118) \\
    3.00  &0.80994 (6)& 3.00344 (101) \\
    2.50  &0.76492 (7)& 2.50242 (85) \\
    2.00  &0.69776 (9)& 2.00308 (69) \\
    1.50  &0.59595 (11)& 1.50129 (55) \\
    1.00  &0.44646 (13)& 1.00114 (43) \\
    0.75  &0.35083 (14)& 0.75063 (39) \\
    0.50  &0.24251 (15)& 0.50052 (35) \\
    0.25  &0.12383 (15)& 0.24988 (32) \\
\hline \hline
\end{tabular}
\caption{\label{tab:2d_u1_l32_l32}{Two-dimensional compact U(1) lattice gauge theory. Summary of the thermodynamic observables: inverse temperature $\beta$, plaquette $P$, and estimated inverse temperature $\beta_M$ on a $32 \times 32$ lattice.}}
	\end{table} 
\endgroup

\begingroup
\renewcommand*{\arraystretch}{1.5}
\begin{table}[ht]
\centering
\begin{tabular}{|p{0.075\textwidth}|p{0.15\textwidth}|p{0.15\textwidth}|p{0.15\textwidth}|}\hline \hline 
~~$\beta$~~ & ~~$P$~~     & ~~$\chi_P$~~ &~~$\beta_M$~~\\ \hline \hline  
4.0000 & 0.93531 (24) & 0.01000 (0) & 4.00026 (816) \\
3.0000 & 0.91262 (32) & 0.01800 (0) & 3.00072 (618) \\
2.0000 & 0.86461 (49) & 0.04100 (1) & 2.00038 (398) \\
1.5000 & 0.81259 (71) & 0.08200 (2) & 1.50075 (307) \\
1.1000 & 0.71596 (108) & 0.18500 (4) & 1.09994 (217) \\
1.0500 & 0.69117 (122) & 0.24100 (5) & 1.04994 (204) \\
1.0300 & 0.67724 (132) & 0.28700 (6) & 1.03003 (209) \\
1.0250 & 0.67289 (132) & 0.29500 (6) & 1.02531 (207) \\
1.0200 & 0.66779 (140) & 0.33100 (8) & 1.02015 (204) \\
1.0150 & 0.66219 (147) & 0.35900 (8) & 1.01496 (203) \\
1.0100 & 0.65489 (173) & 0.47300 (11) & 1.00991 (204) \\
1.0095 & 0.65411 (177) & 0.51000 (13) & 1.00958 (199) \\
1.0090 & 0.65328 (180) & 0.53600 (14) & 1.00950 (200) \\
1.0085 & 0.64424 (140) & 1.49700 (18) & 1.00854 (93) \\
1.0080 & 0.63677 (159) & 2.02600 (17) & 1.00808 (92) \\
1.0075 & 0.62815 (179) & 2.47300 (17) & 1.00759 (94) \\
1.0070 & 0.62606 (171) & 2.33300 (16) & 1.00709 (93) \\
1.0060 & 0.60657 (132) & 1.38800 (17) & 1.00613 (97) \\
1.0050 & 0.60225 (126) & 1.26300 (17) & 1.00512 (96) \\
1.0000 & 0.58524 (196) & 0.62100 (14) & 1.00049 (210) \\
0.9900 & 0.56540 (170) & 0.49300 (12) & 0.99022 (209) \\
0.9500 & 0.51373 (158) & 0.42400 (10) & 0.95021 (209) \\
0.7500 & 0.37296 (160) & 0.43100 (10) & 0.75016 (200) \\
0.5000 & 0.24579 (166) & 0.47200 (10) & 0.50021 (171) \\
0.2500 & 0.12411 (175) & 0.49800 (11) & 0.25004 (155) \\

\hline \hline
\end{tabular}
\caption{\label{tab:4d_u1_l8}{Four-dimensional compact U(1) lattice gauge theory on a $8^4$ lattice. Summary of the thermodynamic observables: inverse temperature $\beta$, plaquette $P$, plaquette susceptibility $\chi_P$, and estimated inverse temperature $\beta_M$.}}
\end{table} 
\endgroup

\raggedright
\bibliographystyle{utphys}
\bibliography{bibfile}
\end{document}